%% file: iclr2027_conference.tex
\documentclass{article}
\usepackage{iclr2027_conference,times}

\input{math_commands.tex}

\usepackage{amsmath,amssymb,amsthm,mathtools}
\usepackage{graphicx}
\usepackage[table]{xcolor}
\usepackage{booktabs}
\usepackage{multirow}
\usepackage{array}
\usepackage{float}
\usepackage{flafter}
\usepackage{wrapfig}
\usepackage{needspace}
\usepackage{tabularx}
\usepackage{algorithm}
\usepackage{algpseudocode}
\usepackage{microtype}
\usepackage{xspace}
\usepackage{url}
\usepackage{hyperref}
\usepackage{etoc}
\usepackage{placeins}
\newcommand{\compactfigurecaptions}{\setlength{\abovecaptionskip}{3pt}\setlength{\belowcaptionskip}{0pt}\setlength{\parskip}{0pt}}
\AddToHook{env/figure/begin}{\compactfigurecaptions}
\AddToHook{env/figure*/begin}{\compactfigurecaptions}
\AddToHook{env/wrapfigure/begin}{\compactfigurecaptions}
\definecolor{venueorange}{HTML}{D97706}
\definecolor{mainresultrow}{HTML}{EFE6D7}
\definecolor{mainimprovrow}{HTML}{F2F3F5}
\newcommand{\venueyear}[1]{\textcolor{venueorange}{\scriptsize #1}}

\hypersetup{
  pdftitle={ResTD: Residual Trajectory Distillation for Generative Retrieval},
  pdfauthor={Weihao Shen, Wei Chen, Fuwei Zhang, Guojun Liu, Qingsong Hua, Wei Lin, Fuzhen Zhuang},
  pdfsubject={arXiv Preprint},
  pdfkeywords={generative retrieval, semantic IDs, residual quantization, process supervision, knowledge distillation},
  colorlinks=true,
  linkcolor=red,
  anchorcolor=red,
  citecolor=brown,
  urlcolor=brown
}

\newtheorem{proposition}{Proposition}
\newcommand{\method}{ResTD\xspace}

\newcolumntype{L}[1]{>{\raggedright\arraybackslash}p{#1}}

\title{Residual Trajectory Distillation for \\ Generative Retrieval}

\author{\textbf{Weihao Shen\textsuperscript{1,2}, Wei Chen\textsuperscript{1,2},
Fuwei Zhang\textsuperscript{1}}\\
\textbf{Guojun Liu\textsuperscript{2},
Qingsong Hua\textsuperscript{2}, Wei Lin\textsuperscript{2},
Fuzhen Zhuang\textsuperscript{1}\thanks{Corresponding author.}}\\[4pt]
{\normalfont\small\textsuperscript{1} Institute of Artificial Intelligence, Beihang University, Beijing, China}\\
{\normalfont\small\textsuperscript{2} Meituan, Beijing, China}\\[2pt]
{\normalfont\small\texttt{\{shenweihao,chenwei23,zhuangfuzhen\}@buaa.edu.cn}}
}

\iclrfinalcopy
\begin{document}
\etocdepthtag.toc{main}
\raggedbottom

\setcounter{footnote}{1}
\maketitle
\lhead{Preprint}

\begin{abstract}

Generative retrieval has emerged as a general retrieval paradigm, representing items with discrete Semantic IDs (SIDs) and retrieving them through autoregressive identifier generation. When SIDs are constructed with residual quantization (RQ), standard retrieval training supervises only the selected codes and discards the residual trajectories that produce them. The same hard code can nevertheless arise from different preferences over competing codewords, while the residual trajectory also contains information about subsequent quantization decisions. As a result, hard SID supervision collapses distinct quantization behaviors into identical targets and leaves information available during indexing unused in retrieval training.
We introduce \textbf{ResTD}, a \textbf{Res}idual \textbf{T}rajectory \textbf{D}istillation framework that transfers this discarded indexing information into retrieval training. Treating the frozen RQ indexer as a process teacher, it distills residual-induced codeword preferences into SID-decoding states. This supervision recovers distinctions hidden by hard assignments and allows earlier decoder states to capture information about subsequent quantization decisions before the corresponding SID suffix is generated. In this way, richer information from SID construction is incorporated into retrieval learning while preserving the original retrieval index and inference procedure.
Experiments on multilingual e-commerce retrieval show consistent improvements over strong baselines and matched training controls. Controlled comparisons show that residual-derived targets outperform the tested codebook-only soft targets. Representation probes further show that future codebook preferences become more recoverable from earlier decoder states. 
ResTD can also be readily extended beyond retrieval to generative recommendation.
 Code is available at: \href{https://github.com/Nevaeh7/iclr2027_ResTD.git}{\textcolor{blue}{\nolinkurl{https://github.com/Nevaeh7/iclr2027_ResTD.git}}}.
\end{abstract}
\section{Introduction}
\label{sec:introduction}

Generative retrieval has emerged as a prominent paradigm that reformulates
retrieval as sequence generation, producing identifiers of
relevant documents or items from a query
\citep{decao2021genre,tay2022transformer,wang2022nci,
sun2023learning,zhang2025merge}.
A central challenge in this paradigm is constructing discrete identifiers
that preserve useful semantic structure.
Learned Semantic IDs (SIDs) have attracted substantial attention, with recent work focusing on their construction and use in generative retrieval
\citep{zeng2024ripor,li2024ltrgr,liu2026catid}.
Most existing generative retrieval methods
\citep{rajput2023recommender,fu2026forge,zhang2026calir}
use the final SID as the supervision target, as it directly serves as the
retrieval address and is generated at inference.
This outcome-level supervision captures only the selected discrete
assignments.
For RQ-based SIDs, the residuals produced during identifier construction
encode finer-grained preferences over competing codewords that are not
reflected in the final SID and thus remain unused in retrieval training.

\begin{figure}[!t]
  \centering
  \includegraphics[width=\linewidth]{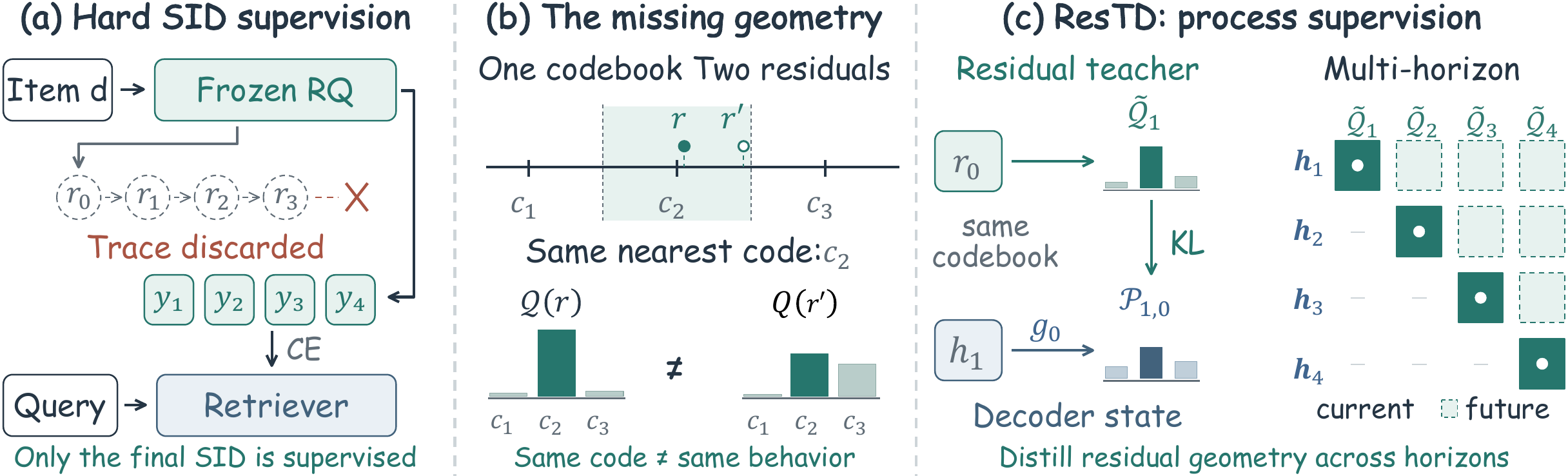}
  \caption{ResTD uses the frozen RQ indexer as a process teacher, distilling
  current and future codebook preferences into SID-decoding states. The
  retrieval index and inference remain unchanged.}
  \vspace{-1em}
  \label{fig:overview}
\end{figure}

This information loss is particularly pronounced for residual-quantized
identifiers.
Residual quantization (RQ) constructs an SID progressively by selecting a
codeword at each level and passing the remaining residual to the next
\citep{gray1984vector,oord2017vqvae,lee2022rqvae,
rajput2023recommender}.
The selected code alone does not fully characterize the residual that produced
it.
As illustrated in Figure~\ref{fig:overview}, two residuals may induce different
preferences over competing codewords while sharing the same selected code,
making them indistinguishable under hard token supervision.
Their relative distances and margins to alternative codewords can differ
substantially within the same quantization cell.
These distinctions also extend beyond the current level, since the remaining
residual determines the input to subsequent quantization stages and influences
later code selections.
The trajectory therefore retains information about current preferences and
subsequent quantization decisions.
Hard SID supervision provides no signal to preserve this information in
decoder states.
Consequently, the final SID records the quantization outcomes but omits the
fine-grained structure underlying their formation, creating an information gap
between indexing and retrieval training.

Residual-distance distributions provide a natural way to recover information
otherwise discarded by hard SID supervision
\citep{hinton2015distilling,lee2022rqvae}.
Yet using them for retrieval is not straightforward.
An SID is both a quantization outcome and an address in a fixed retrieval
index.
The preferences induced by residual geometry therefore need not align with
the retrieval target.
A nearby codeword may correspond to another item or an invalid path, and
quantization proximity itself does not imply query relevance.
Moreover, collision resolution may cause the stored SID to differ from the
code preferred by the reconstructed residual.
Directly distilling these preferences can therefore conflict with the
discrete target used for retrieval.
Residual supervision must preserve the informative structure exposed during
quantization while remaining consistent with the stored retrieval address.

To satisfy these requirements, we introduce ResTD, a residual trajectory
distillation framework that treats the frozen RQ indexer as a process teacher.
ResTD transfers residual-induced codeword preferences into query-conditioned
SID-decoding states, recovering distinctions obscured by hard assignments while
remaining consistent with the stored SID.
This consistency preserves the retrieval address and incorporates richer
information from SID construction into retrieval learning without altering the
original identifier space.
Beyond the current quantization decision, ResTD further supervises earlier
decoder states with preferences from subsequent quantization levels across
multiple future horizons before the corresponding SID suffix enters the
autoregressive context.
Prior work has used future information to improve sequence representations
through future-state matching or multi-token prediction
\citep{serdyuk2018twin,qi2020prophetnet,gloeckle2024better}.
Here, item-conditioned codeword distributions serve as future targets that
retain distinctions not expressed by future hard tokens alone.
This encourages earlier decoder states to encode structure associated with the
remaining identifier before those codes are generated, providing richer
supervision for SID prediction.

We evaluate ResTD on the ESCI product-retrieval benchmark
\citep{reddy2022esci} in English, Spanish, and Japanese.
ResTD consistently outperforms strong baselines across all 15 reported
evaluation settings. Controlled comparisons show benefits over future
hard-target supervision, codebook-only targets, and shape-matched teachers
from other items. Representation probes further show that future codebook
preferences are more recoverable from earlier decoder states. We also extend
ResTD to generative recommendation, where it consistently improves performance
across multiple datasets.

Our contributions are threefold:

\noindent\textbf{(i)}
We identify a gap between RQ-based SID construction and retrieval training,
where hard SID supervision discards residual structure associated with current
and subsequent quantization decisions.

\noindent\textbf{(ii)}
We propose ResTD, which turns a frozen RQ indexer into a process
teacher and transfers current and future residual-induced preferences from the
indexing process into SID-decoding states while preserving the deployed
retrieval addresses and autoregressive inference procedure.

\noindent\textbf{(iii)}
ResTD improves retrieval and recommendation.
Controls assess residual structure and correspondence between items and teachers;
probes examine future preferences in early decoder states.

\section{Related Work}
\label{sec:related_work}
\textbf{Generative retrieval and learned identifiers.}
Generative retrieval autoregressively produces entity names, document spans,
or discrete identifiers
\citep{decao2021genre,bevilacqua2022sealm,tay2022transformer,wang2022nci}.
A major line of work develops identifiers preserving document or item
semantics. GenRet learns discrete document tokenization
\citep{sun2023learning}, while TIGER constructs Semantic IDs with residual
quantization \citep{rajput2023recommender}. LETTER and MERGE incorporate
collaborative and relevance signals into identifier learning
\citep{wang2024learnable,zhang2025merge}. Beyond identifier construction,
LTRGR introduces ranking-aware training \citep{li2024ltrgr}, while PAG, CaLIR,
and PRO advance autoregressive search, latent intent modeling, and prefix
retention, respectively
\citep{zeng2024pag,zhang2026calir,chen2026pro}. ResTD complements these directions by deriving supervision from residual
quantization during the construction of a fixed RQ-based index, using
additional indexing information that is discarded once the final SID is formed.

\textbf{Distillation and codebook supervision.}
Knowledge distillation transfers information beyond hard labels through soft
predictions, intermediate representations, generated sequences, or structural
relations
\citep{hinton2015distilling,kim2016sequence,romero2015fitnets,
park2019relational}. In quantized generative models, RQ-Transformer derives
soft code targets from residual distances \citep{lee2022rqvae}. ETEGRec aligns
sequence representations with item code distributions and reconstructed item
embeddings \citep{liu2025etegrec}, while TARQ aligns residual-to-code
distributions for target-attention approximation \citep{li2026tarq}. SmartGR
distills hierarchical SID predictions and beam-ranking preferences from a
stronger recommender \citep{zhang2026smartgr}. ResTD instead treats the
frozen RQ indexer itself as the teacher: residual preferences are reconstructed
along stored SID paths and directly distilled into SID-decoding states without
changing the underlying identifier space.

\textbf{Future supervision.}
Future information has long been used to strengthen earlier sequence
representations. Twin Networks regularize forward states using future context
\citep{serdyuk2018twin}; ProphetNet and multi-token prediction supervise a
state with multiple subsequent tokens
\citep{qi2020prophetnet,gloeckle2024better}; and recent work further explores
compact future summaries as predictive targets
\citep{mahajan2026beyond}. ResTD follows this line of work by using residual
codebook distributions as structured future targets.
Earlier decoder states are thereby explicitly supervised by both future SID
assignments and codeword preferences at subsequent quantization levels.

\begin{figure}[!t]
  \centering
  \includegraphics[width=\linewidth]{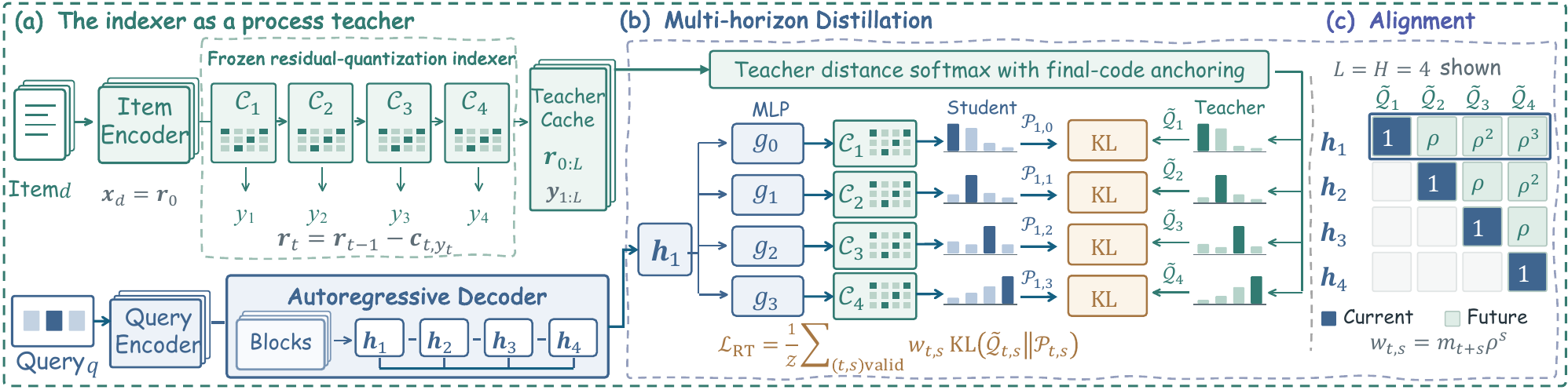}
  \caption{Overview of ResTD. The frozen RQ indexer supplies residual-derived
  teacher distributions along stored SID paths; horizon-specific heads align decoder states
  with current and future targets.}
  \label{fig:method}
  \vspace{-1em}
\end{figure}

\section{Preliminaries: Residual-Quantized Retrieval}
\label{sec:prelim}

\begingroup
\clubpenalty=10000
\widowpenalty=10000
\displaywidowpenalty=10000

\textbf{Residual-quantized indexing.}
Residual quantization (RQ) represents an item through successive vector
quantization over level-specific codebooks
\citep{oord2017vqvae,lee2022rqvae,rajput2023recommender}.
Let $\vx_d\in\mathbb{R}^{D}$ denote the representation of item $d$, and let
$\mathcal{C}_t=\{\vc_{t,j}\}_{j=1}^{K_t}$ be the codebook at level $t$.
Starting from $\vr_0=\vx_d$, the $t$-th quantization level selects the nearest
codeword and updates the residual for the next level as follows:
\begin{equation}
  y_t
  =
  \argmin_{j}
  \|\vr_{t-1}-\vc_{t,j}\|_2^2,
  \qquad
  \vr_t
  =
  \vr_{t-1}-\vc_{t,y_t}.
  \label{eq:rq}
\end{equation}
After $L$ levels, the indexer yields the Semantic ID
$\vy(d)=(y_1,\ldots,y_L)$ and the residual trajectory
$(\vr_0,\ldots,\vr_L)$. The SID records the discrete quantization path, while
the residual trajectory retains the intermediate states that determine each
codeword selection.

\textbf{Autoregressive SID retrieval.}
Given a query $q$, a generative retriever with parameters $\theta$ predicts the
target SID autoregressively. Under teacher forcing, we use the standard
sequence-generation objective for SID prediction
\citep{sutskever2014sequence,tay2022transformer}
\begin{equation}
  \mathcal{L}_{\mathrm{SID}}
  =
  -\sum_{t=1}^{L}
  \log p_\theta(y_t\mid q,y_{<t}).
  \label{eq:sidce}
\end{equation}
Let $\vh_t$ denote the decoder state used to predict $y_t$. In CaLIR
\citep{zhang2026calir}, $\vh_1$ corresponds to the final reasoning state and
$\vh_{2:L}$ to subsequent SID-decoding states. \method applies auxiliary
supervision only to states that directly predict SID tokens.

\textbf{Hard assignments and residual geometry.}
Equation~\ref{eq:rq} retains only the nearest-code assignment. Within the same
quantization cell, however, residuals can have different distances and margins
to the remaining codewords \citep{gray1984vector}. Consequently, identical
hard SID targets need not imply identical local codebook preferences. This
distinction between the selected code and the residual geometry underlying the
selection forms the basis of the supervision introduced next.

\endgroup
\section{ResTD: Learning from the Indexing Process}
\label{sec:method}

ResTD uses the frozen RQ indexer as a structured teacher in the codebook
geometry that defines the Semantic IDs. The supervision retains the
residual-induced preferences available before each discrete code selection,
recovering information that the selected code itself cannot express. The frozen
codebooks provide a shared reference space: indexer residuals define teacher
distributions over codewords, and projected decoder states define student
distributions in the same space. We first align the two at the current
quantization level and then extend the supervision across the remaining
residual trajectory. Since the retrieval index remains fixed, each teacher
target must stay consistent with the stored SID without altering the deployed
identifier space. Figure~\ref{fig:method} summarizes the framework.

\subsection{Residual Behavioral Distillation}
\label{sec:rbd}

At quantization level $t$, the residual $\vr_{t-1}$ induces a complete set of
distances to the codewords in $\mathcal{C}_t$. Standard RQ compresses this
geometry into the nearest-code assignment in Equation~\ref{eq:rq}. ResTD
instead retains the pre-assignment geometry through the teacher distribution
\begin{equation}
  \gQ_t(j\mid d)
  =
  \frac{
    \exp\!\left(
      -\|\vr_{t-1}-\vc_{t,j}\|_2^2/\tau_T
    \right)
  }{
    \sum_{m=1}^{K_t}
    \exp\!\left(
      -\|\vr_{t-1}-\vc_{t,m}\|_2^2/\tau_T
    \right)
  },
  \label{eq:teacher}
\end{equation}
where $\tau_T$ controls the concentration of the distribution. Unlike the hard
assignment $y_t$, $\gQ_t$ retains both the ordering and relative margins among
competing codewords. In particular, for any pair $j,k$,
\begin{equation}
  \log
  \frac{\gQ_t(j\mid d)}{\gQ_t(k\mid d)}
  =
  \frac{
    2\vr_{t-1}^{\top}(\vc_{t,j}-\vc_{t,k})
    +
    \|\vc_{t,k}\|_2^2
    -
    \|\vc_{t,j}\|_2^2
  }{\tau_T}.
  \label{eq:teacher-logodds}
\end{equation}
Thus, two residuals assigned to the same hard code can induce different
pairwise preferences over the remaining codebook. The teacher distribution
captures this item-side quantization geometry; query relevance remains
supervised by the original retrieval objective.

Let $\vh_t$ denote the decoder state used to predict $y_t$. We introduce a
learnable projection $g_0:\mathbb{R}^{d_h}\rightarrow\mathbb{R}^{D}$ and
score the projected state against the frozen codebook at level $t$:
\begin{equation}
  \gP_{t,0}(j\mid q,d)
  =
  \softmax_j\!\left(
    -\frac{
      \|g_0(\vh_t)-\vc_{t,j}\|_2^2
    }{\tau_S}
  \right),
  \label{eq:student}
\end{equation}
where $\tau_S$ is the student temperature. Teacher and student
distributions are induced by distances to $\mathcal{C}_t$ and therefore share
the codebook geometry. This permits comparison within the quantization space
used to construct the SID.
Residual Behavioral Distillation (RBD) minimizes the average teacher--student
divergence in the frozen codebook space:
\begin{equation}
  \mathcal{L}_{\mathrm{RBD}}
  =
  \frac{1}{L}
  \sum_{t=1}^{L}
  \KL\!\left(
    \gQ_t
    \,\|\,
    \gP_{t,0}
  \right).
  \label{eq:rbd}
\end{equation}
The resulting objective also admits a geometric interpretation. For a fixed
teacher, its gradient with respect to the projected state is
\begin{equation}
  \nabla_{g_0(\vh_t)}
  \KL\!\left(\gQ_t\,\|\,\gP_{t,0}\right)
  =
  \frac{2}{\tau_S}
  \left(
    \sum_{j} \gP_{t,0}(j\mid q,d)\vc_{t,j}
    -
    \sum_{j} \gQ_t(j\mid d)\vc_{t,j}
  \right).
  \label{eq:rbd-geometry}
\end{equation}
RBD therefore aligns the codebook-weighted expectations induced by the decoder
state and the indexer residual. In practice, $\gQ_t$ is replaced by the
collision-aware teacher introduced in Section~\ref{sec:collision}, and only
valid SID positions contribute to the loss. The original SID objective remains
unchanged, so the retriever is jointly supervised by the target identifier and
the geometry underlying its construction.

\subsection{Multi-Horizon Residual Trajectory Distillation}
\label{sec:multihorizon}

Current-level RBD transfers the quantization preference associated with the
code currently being generated. The residual trajectory also captures the subsequent decisions required to complete the SID.
ResTD uses these subsequent preferences to supervise earlier decoder states
before the corresponding suffix codes enter the autoregressive context.
We introduce $H$ horizon-specific projections
$\{g_s\}_{s=0}^{H-1}$. For state $\vh_t$ and horizon offset $s$, the target
quantization level is $t+s$, with residual state $\vr_{t+s-1}$ and codebook
$\mathcal{C}_{t+s}$. The corresponding teacher and student distributions are
defined as follows:
\begin{align}
  \gQ_{t,s}(j\mid d)
  &=
  \softmax_j\!\left(
    -\frac{
      \|\vr_{t+s-1}-\vc_{t+s,j}\|_2^2
    }{\tau_T}
  \right),
  \label{eq:mhteacher}
  \\
  \gP_{t,s}(j\mid q,d)
  &=
  \softmax_j\!\left(
    -\frac{
      \|g_s(\vh_t)-\vc_{t+s,j}\|_2^2
    }{\tau_S}
  \right).
  \label{eq:mhstudent}
\end{align}

For position $t$, valid offsets satisfy $0\leq s<H_t$, where
$H_t=\min(H,L-t+1)$. The case $s=0$ recovers current-level RBD, while
$s>0$ associates the same decoder state with preferences from later
quantization levels. The resulting supervision is triangular: early decoder
states are constrained by a longer suffix of the residual trajectory, whereas
the number of valid future targets decreases toward the end of the SID.
We aggregate all valid state--horizon pairs using the normalized objective:
\begin{equation}
  \mathcal{L}_{\mathrm{RT}}
  =
  \frac{1}{\gZ}
  \sum_{t=1}^{L}
  \sum_{s=0}^{H_t-1}
  m_{t+s}\rho^s
  \KL\!\left(
    \widetilde{\gQ}_{t,s}
    \,\|\,
    \gP_{t,s}
  \right),
  \qquad
  \gZ
  =
  \sum_{t=1}^{L}
  \sum_{s=0}^{H_t-1}
  m_{t+s}\rho^s ,
  \label{eq:retrace}
\end{equation}
where $m_{t+s}$ masks invalid target positions and $0<\rho\leq1$ discounts
increasingly distant horizons. For each offset $s$, the normalized trajectory
mass used by the objective is
\begin{equation}
  w_s
  =
  \frac{
    \rho^s\sum_{t=1}^{L-s}m_{t+s}
  }{\gZ},
  \qquad
  \sum_{s=0}^{H-1} w_s = 1.
  \label{eq:horizon-mass}
\end{equation}
When all positions are valid, this reduces to
$w_s=(L-s)\rho^s/\sum_{u=0}^{H-1}(L-u)\rho^u$.
The horizon therefore determines both the range and relative weighting of
future supervision. When $H=1$, Equation~\ref{eq:retrace} reduces to
current-level RBD. The target $\widetilde{\gQ}_{t,s}$ preserves the relative
preferences among competing codewords in addition to the selected future code.
These soft targets separate items that share the same future assignment and
encourage earlier decoder states to encode future codebook preferences before
the corresponding SID suffix is generated.

\subsection{Collision-Aware Teacher Construction}
\label{sec:collision}

Residual supervision must remain consistent with the SID stored in the
retrieval index. Collision resolution can modify one or more nearest-code
assignments to avoid duplicate identifiers, so a residual trajectory computed
from the original RQ decisions need not follow the final retrieval path.
We therefore reconstruct the trajectory along the stored SID. Given
$\vy^{\mathrm{final}}
=(y_1^{\mathrm{final}},\ldots,y_L^{\mathrm{final}})$,
we initialize $\vr_0=\vx_d$ and recursively compute
$\vr_t=\vr_{t-1}-\vc_{t,y_t^{\mathrm{final}}}$.
The resulting residual states are consequently defined on the same discrete
path as the retrieval target.
Path reconstruction alone does not guarantee agreement between residual
geometry and the stored code. At a given target, let $q_y$ denote the
uncorrected probability of the stored code and $q_{\max}$ the largest
probability among competing codes. We seek the smallest interpolation toward
the stored assignment that enforces the target margin $\delta$ for each
state--horizon pair:
\begin{equation}
  \epsilon_{t,s}^{\star}
  =
  \min_{\alpha\in[0,1]} \alpha
  \quad
  \mathrm{s.t.}
  \quad
  (1-\alpha)q_y+\alpha
  \geq
  (1-\alpha)q_{\max}+\delta.
  \label{eq:collision-minimal}
\end{equation}
For each state--horizon pair, the corrected teacher distribution over the
target codebook is
\begin{equation}
  \widetilde{\gQ}_{t,s}(j)
  =
  (1-\epsilon_{t,s})\gQ_{t,s}(j)
  +
  \epsilon_{t,s}\,
  \mathbf{1}\!\left[
    j=y_{t+s}^{\mathrm{final}}
  \right].
  \label{eq:collision}
\end{equation}
Solving Equation~\ref{eq:collision-minimal} gives the closed-form correction required
to enforce the stored-code margin:
\begin{equation}
  \epsilon_{t,s}
  =
  \max\!\left\{
    \epsilon,\,
    \left[
      \frac{q_{\max}-q_y+\delta}
           {1-q_y+q_{\max}}
    \right]_0^1
  \right\},
  \label{eq:collision_eps}
\end{equation}
where $[\cdot]_0^1$ clips to $[0,1]$ and $\epsilon$ is the minimum mixing
weight. The correction is therefore the minimum adaptive perturbation, subject
to the floor $\epsilon$, required to make the stored code preferred by the
teacher. Since every non-target probability is multiplied by the same factor
$1-\epsilon_{t,s}$, their pairwise ratios remain unchanged whenever
$\epsilon_{t,s}<1$. The procedure thus restores consistency with the fixed SID
while preserving the residual-induced preference structure among alternatives.

\input{tables/table_rq1_overall}

\subsection{Training and Inference}
\label{sec:training}

ResTD augments the original generative retrieval objective with residual
trajectory distillation:
\begin{equation}
  \mathcal{L}
  =
  \mathcal{L}_{\mathrm{GR}}
  +
  \lambda(u)\mathcal{L}_{\mathrm{RT}},
  \qquad
  \lambda(u)
  =
  \lambda_{\max}
  \min\!\left(
    1,
    \frac{u}{u_{\mathrm{warm}}}
  \right),
  \label{eq:total}
\end{equation}
where $u$ is the optimization step and $u_{\mathrm{warm}}$ controls a linear
warmup of the trajectory objective. In our CaLIR instantiation,
$\mathcal{L}_{\mathrm{GR}}$ is unchanged and contains the original
SID-generation and latent-reasoning objectives. The RQ indexer and codebooks
remain fixed throughout training, while the retriever and horizon-specific
projections are optimized jointly.
During continuation training, the contribution of the trajectory objective is
bounded relative to the SID objective so that residual supervision remains
auxiliary to the retrieval target. At inference, the residual teachers and
projection functions are removed. Retrieval therefore uses the original
SID-generation distribution and trie-constrained autoregressive search, with
no modification to the retrieval index or inference procedure.

\section{Experiments}
\label{sec:experiments}
We evaluate ResTD through seven research questions:
\textbf{RQ1}: How does the method compare with sparse, dense, and
generative retrieval approaches?
\textbf{RQ2}: How do residual supervision, multi-horizon distillation, and
collision-aware teachers affect retrieval under matched settings?
\textbf{RQ3}: Which teacher target properties contribute to retrieval gains?
\textbf{RQ4}: Do learned decoder states encode current and future codebook
preferences from the indexer?
\textbf{RQ5}: How do teacher disagreement and relevant-item counts relate to
retrieval gains?
\textbf{RQ6}: Does residual trajectory supervision generalize to generative
recommendation?
\textbf{RQ7}: How can the advantage of ResTD be understood intuitively?
\subsection{Experimental Setup}
\label{sec:exp-setup}

\noindent\textbf{(1) Datasets.}
We evaluate \method on ESCI-US, ESCI-ES, and ESCI-JP
(English, Spanish, and Japanese) from the Shopping Queries Dataset (ESCI),
a large-scale multilingual product retrieval benchmark
\citep{reddy2022esci}, following the filtering and splits of
\citet{zhang2026calir}.
\noindent\textbf{(2) Baselines.}
We compare against sparse (BM25), dense (DPR, MPNet, Sentence-T5,
Sentence-mT5, BGE-M3, and LaSER), and generative retrievers
(DSI, TIGER, Hi-Gen, LTRGR, RIPOR, CAT-ID$^2$, MERGE, FORGE,
HGRec, and CaLIR). The overall comparison includes published results
under their original training protocols. Controlled comparisons start
from the same reproduced CaLIR model under matched training and
evaluation settings: continued fine-tuning (Base-FT), current-level RBD
($H=1$), multi-horizon ResTD, future hard-SID supervision,
shuffled-teacher variants, full-output distillation, and non-target
output alignment.
\noindent\textbf{(3) Implementation Details.}
We use T5-base for ESCI-US and mT5-base for ESCI-ES/JP.
The SID index has four quantization levels with 256 codewords each.
Unless otherwise specified, $H=4$, $\lambda_{\max}=0.1$,
$\tau_T=\tau_S=0.2$, $\rho=0.7$, and the collision-correction floor
$\epsilon=0.1$. Each controlled continuation uses 1,200 updates,
effective batch size 128, learning rate $10^{-5}$, and paired seeds
42, 2027, and 2028. Retrieval uses trie-constrained beam search
with beam size 100. We report Recall@5/10/100 and NDCG@10/100
on the percentage scale. Further experimental details and hyperparameter
sensitivity appear in Appendices~\ref{app:experimental-setup}
and~\ref{app:hyperparameter-sensitivity}.

\subsection{Experimental Results and Discussion}
\label{sec:exp-results}

\begingroup
\setlength{\emergencystretch}{1.5em}

\phantomsection
\textbf{Main Experimental Results (RQ1).}
\label{sec:main-results}
Table~\ref{tab:main} shows that our method achieves the best results on
every reported metric across ESCI-US, ESCI-ES, and ESCI-JP, consistently
outperforming sparse, dense, and generative retrieval baselines.
Our advantage is most pronounced at early ranks. Higher Recall and NDCG
at these cutoffs show that ResTD brings more relevant items to the top
of the retrieved list and improves their ordering. Recall@100 also
increases in every dataset, indicating that our method combines stronger
early ranking with broader coverage of relevant items. The gains over
CaLIR, the strongest baseline, demonstrate the value of our residual
trajectory supervision for generative retrieval. Our method enriches SID
prediction with preferences over competing codewords, transferring
information from identifier construction into query-conditioned retrieval
learning.
Our method consistently delivers these gains in English, Spanish, and
Japanese using the same distillation hyperparameters. ResTD turns residual
information from SID construction into more effective query-to-identifier
prediction. Dataset preparation, baseline
settings, relevance judgments, and the exact SID-level evaluation protocol for
multilingual product retrieval are detailed in
Appendix~\ref{app:experimental-setup}.

\phantomsection
\textbf{Matched Training Objectives (RQ2).}
\label{sec:training-comparisons}Table~\ref{tab:core-training}
\input{tables/table_core_training}
compares our supervision objectives across three paired seeds under the
same training budget. Current-level RBD ($H=1$) improves every reported
metric over Base-FT and outperforms full-output KD with the same teacher,
demonstrating the benefit of distilling residual preferences into decoder
states. Extending supervision to $H=2$ and then $H=4$ yields progressive
gains across all metrics and locales. Our multi-horizon formulation thus
benefits from information at increasingly distant quantization levels.
ResTD also surpasses RBD + future hard SID under matched horizon and
supervision weights, improving the ranking and coverage of relevant items.
This comparison establishes the value of future codeword preferences
beyond future hard labels. Shuffled teachers yield retrieval scores close
to Base-FT in all three locales. Collision-aware correction further improves
Recall@100 and NDCG@10 in every locale by aligning residual preferences
with the stored identifiers. The gains persist across moderate correction
floors, with the default attaining the highest observed means.
Appendix~\ref{app:output-supervision} defines the output-space controls,
and Appendix~\ref{app:collision-correction-sensitivity} details the
correction rule and sensitivity to the correction strength.
\begin{figure}[t]
  \centering
  \begin{minipage}[t]{0.485\textwidth}
    \centering
    \begin{minipage}[c][0.62\linewidth][c]{\linewidth}
      \centering
      \includegraphics[width=\linewidth]{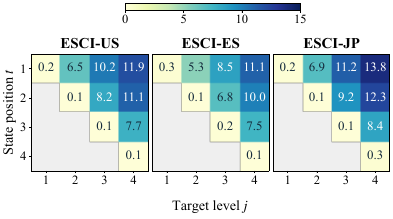}
    \end{minipage}
    \caption{Probe KL reduction (\%) for ResTD ($H=4$) over RBD ($H=1$).
    Diagonal cells are current targets; rightward cells are future targets.}
    \vspace{-0.8em}
    \label{fig:future-probe}
  \end{minipage}\hfill
  \begin{minipage}[t]{0.485\textwidth}
    \centering
    \begin{minipage}[c][0.62\linewidth][c]{\linewidth}
      \centering
      \includegraphics[width=\linewidth]{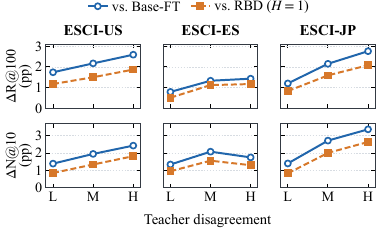}
    \end{minipage}
    \caption{Mean ResTD gains over Base-FT and RBD ($H=1$), averaged across
    three seeds. L/M/H denote increasing teacher disagreement.}
    \vspace{-0.8em}
    \label{fig:disagreement-gains}
  \end{minipage}
\end{figure}
\input{sections/teacher_geometry_analysis}

\phantomsection
\textbf{Representation Alignment (RQ4).}
\label{sec:representation-transfer}\input{sections/future_probe_analysis}

\input{sections/multipositive_analysis}

\input{sections/generative_recommendation}

\input{sections/case_study_analysis}

\endgroup

\section{Conclusion}

We introduced ResTD, a residual trajectory distillation framework that
transfers current and future codebook preferences from a frozen RQ indexer
into query-conditioned decoder states, enriching hard SID supervision
and preserving the index and autoregressive inference.
Across three ESCI locales, ResTD consistently outperforms continued
training, current-level RBD, and future hard-SID supervision under
matched settings. Controlled comparisons and representation probes
support the value of residual-derived targets and improved access to
future codebook information from early decoder states.
Results on Beauty, Instruments, and Yelp extend ResTD to generative
recommendation.

\subsection*{AI use statement}

We used generative AI solely as a writing aid to refine the wording and improve
the readability of the manuscript. We did not use it to formulate the research
questions, design the methodology or experiments, or analyze and interpret the
results. We reviewed all AI-assisted revisions and take full responsibility for
the final version of the paper.

\subsection*{Ethics statement}

We have reviewed the potential ethical implications of this work and do not
identify any particular ethical issues that warrant additional discussion.

\subsection*{Reproducibility statement}

To facilitate reproducibility, we provide detailed descriptions of the proposed
method, experimental configuration, and evaluation procedure in the main paper
and appendix. The source code and relevant implementation details are also made
available through our GitHub repository to support independent reproduction
of the reported results.

\bibliography{iclr2027_conference}
\bibliographystyle{iclr2027_conference}

\clearpage
\appendix
\input{sections/appendix_layout}
\etocdepthtag.toc{appendix}
\begingroup
  \hypersetup{linkcolor=black}
  \etocsettagdepth{main}{none}
  \etocsettagdepth{appendix}{subsection}
  \etocsetnexttocdepth{subsection}
  \etocsettocstyle{\begin{center}\Large\bfseries Appendix\end{center}
    \noindent Experimental protocols, derivations, and additional analyses
    for ResTD.\par\medskip
    \noindent\textbf{Contents}\par\smallskip
    \small
  }{\par\medskip}
  \tableofcontents
\endgroup

\input{sections/appendix_experimental_setup}
\FloatBarrier

\section{Method and Training Details}
\label{app:method-details}
\input{sections/appendix_geometry_details}
\input{sections/appendix_algorithm}
\input{sections/appendix_computational_cost}

\FloatBarrier

\section{Supplementary Experiments and Analyses}
\label{app:supplementary-experiments}
We examine generalization to NQ320K, teacher targets, output-space
supervision, representation probes, multi-positive queries, collision
correction, and hyperparameter sensitivity.

\input{sections/appendix_nq320k}
\FloatBarrier
\input{sections/appendix_teacher_geometry}
\FloatBarrier
\input{sections/appendix_output_supervision}
\FloatBarrier
\input{sections/appendix_future_probe}
\FloatBarrier
\input{sections/appendix_multipositive}
\input{sections/appendix_collision_teacher_analysis}
\FloatBarrier
\input{sections/appendix_hyperparameter_sensitivity}
\FloatBarrier

\input{sections/appendix_case_study}

\end{document}

%% file: math_commands.tex
\usepackage{amsmath,amsfonts,bm}

\def\eqref#1{equation~\ref{#1}}

\def\1{\bm{1}}

\def\vc{{\bm{c}}}

\def\vh{{\bm{h}}}

\def\vr{{\bm{r}}}
\def\vs{{\bm{s}}}

\def\vu{{\bm{u}}}

\def\vx{{\bm{x}}}
\def\vy{{\bm{y}}}

\DeclareMathAlphabet{\mathsfit}{\encodingdefault}{\sfdefault}{m}{sl}
\SetMathAlphabet{\mathsfit}{bold}{\encodingdefault}{\sfdefault}{bx}{n}

\def\gP{{\mathcal{P}}}
\def\gQ{{\mathcal{Q}}}

\def\gZ{{\mathcal{Z}}}

\newcommand{\softmax}{\mathrm{softmax}}

\newcommand{\KL}{D_{\mathrm{KL}}}

\DeclareMathOperator*{\argmin}{arg\,min}

%% file: tables/table_rq1_overall.tex
\begin{table}[!t]
\caption{Overall ESCI retrieval results (\%) across sparse, dense, and
generative methods. ResTD reports three-seed means. Relative improvements
compare with the strongest baseline in each column.}
\label{tab:main}
\centering
\newsavebox{\overallnumberbox}
\newcommand{\overallnumber}[2][\relax]{\sbox{\overallnumberbox}{#1{#2}}\makebox[\wd\overallnumberbox][r]{\raisebox{0pt}[\ht\overallnumberbox][\dp\overallnumberbox]{#1{{\fontsize{10.5}{11}\selectfont #2}}}}}
\resizebox{\textwidth}{!}{\begin{tabular}{lrrrrrrrrrrrrrrr}
\toprule
& \multicolumn{5}{c}{ESCI-US} & \multicolumn{5}{c}{ESCI-ES} & \multicolumn{5}{c}{ESCI-JP}\\
\cmidrule(lr){2-6}\cmidrule(lr){7-11}\cmidrule(lr){12-16}
Model & R@5 & R@10 & R@100 & N@10 & N@100
& R@5 & R@10 & R@100 & N@10 & N@100
& R@5 & R@10 & R@100 & N@10 & N@100\\
\midrule
\multicolumn{16}{l}{\emph{Sparse retrieval}}\\
BM25 \venueyear{(FnT IR'09)} &\overallnumber{4.02}&\overallnumber{5.84}&\overallnumber{14.08}&\overallnumber{5.64}&\overallnumber{8.11} &\overallnumber{4.13}&\overallnumber{6.15}&\overallnumber{16.21}&\overallnumber{7.45}&\overallnumber{10.20} &\overallnumber{4.70}&\overallnumber{7.96}&\overallnumber{16.73}&\overallnumber{8.63}&\overallnumber{11.25}\\
\midrule
\multicolumn{16}{l}{\emph{Dense retrieval}}\\
DPR \venueyear{(EMNLP'20)} &\overallnumber{5.54}&\overallnumber{8.93}&\overallnumber{29.30}&\overallnumber{8.17}&\overallnumber{15.55} &\overallnumber{5.24}&\overallnumber{7.08}&\overallnumber{25.27}&\overallnumber{8.52}&\overallnumber{14.33} &\overallnumber{4.18}&\overallnumber{7.26}&\overallnumber{23.84}&\overallnumber{8.98}&\overallnumber{14.73}\\
MPNet \venueyear{(NeurIPS'20)} &\overallnumber{2.76}&\overallnumber{4.58}&\overallnumber{15.30}&\overallnumber{4.12}&\overallnumber{9.83} &\overallnumber{2.53}&\overallnumber{4.02}&\overallnumber{13.27}&\overallnumber{5.87}&\overallnumber{8.91} &\overallnumber{1.13}&\overallnumber{1.97}&\overallnumber{5.58}&\overallnumber{1.79}&\overallnumber{2.82}\\
Sentence-T5 \venueyear{(ACL'22)} &\overallnumber{4.63}&\overallnumber{6.97}&\overallnumber{24.59}&\overallnumber{6.84}&\overallnumber{11.58} &--&--&--&--&-- &--&--&--&--&--\\
Sentence-mT5 \venueyear{(COLING'24)} &--&--&--&--&-- &\overallnumber{5.44}&\overallnumber{8.68}&\overallnumber{30.04}&\overallnumber{8.42}&\overallnumber{15.09} &\overallnumber{5.69}&\overallnumber{8.73}&\overallnumber{29.01}&\overallnumber{9.19}&\overallnumber{15.70}\\
BGE-M3 \venueyear{(ACL'24)} &\overallnumber{6.59}&\overallnumber{9.82}&\overallnumber{32.29}&\overallnumber{9.91}&\overallnumber{16.77} &\overallnumber{5.71}&\overallnumber{9.07}&\overallnumber{28.75}&\overallnumber{9.76}&\overallnumber{16.02} &\overallnumber{5.14}&\overallnumber{8.92}&\overallnumber{27.93}&\overallnumber{10.12}&\overallnumber{17.02}\\
LaSER \venueyear{(SIGIR'26)} &\overallnumber{6.52}&\overallnumber{10.23}&\overallnumber{35.37}&\overallnumber{8.19}&\overallnumber{15.74} &\overallnumber{5.28}&\overallnumber{8.40}&\overallnumber{32.22}&\overallnumber{8.69}&\overallnumber{16.16} &\overallnumber{5.88}&\overallnumber{8.78}&\overallnumber{28.66}&\overallnumber{9.28}&\overallnumber{15.53}\\
\midrule
\multicolumn{16}{l}{\emph{Generative retrieval}}\\
DSI\textsubscript{naive} \venueyear{(NeurIPS'22)} &\overallnumber{0.42}&\overallnumber{1.74}&\overallnumber{2.03}&\overallnumber{0.32}&\overallnumber{0.99} &\overallnumber{0.28}&\overallnumber{0.94}&\overallnumber{1.83}&\overallnumber{0.77}&\overallnumber{1.60} &\overallnumber{0.19}&\overallnumber{0.24}&\overallnumber{1.35}&\overallnumber{0.21}&\overallnumber{1.11}\\
DSI\textsubscript{semantic} \venueyear{(NeurIPS'22)} &\overallnumber{3.74}&\overallnumber{6.02}&\overallnumber{20.69}&\overallnumber{6.24}&\overallnumber{10.60} &\overallnumber{4.27}&\overallnumber{7.13}&\overallnumber{21.63}&\overallnumber{9.92}&\overallnumber{14.23} &\overallnumber{4.08}&\overallnumber{7.40}&\overallnumber{23.90}&\overallnumber{9.15}&\overallnumber{14.58}\\
TIGER \venueyear{(NeurIPS'23)} &\overallnumber{4.79}&\overallnumber{7.84}&\overallnumber{25.98}&\overallnumber{7.24}&\overallnumber{12.68} &\overallnumber{3.83}&\overallnumber{7.31}&\overallnumber{25.03}&\overallnumber{8.72}&\overallnumber{14.51} &\overallnumber{4.67}&\overallnumber{7.98}&\overallnumber{25.89}&\overallnumber{9.67}&\overallnumber{15.52}\\
Hi-Gen \venueyear{(ICDM'24)} &\overallnumber{3.13}&\overallnumber{4.97}&\overallnumber{15.91}&\overallnumber{5.25}&\overallnumber{8.55} &\overallnumber{2.97}&\overallnumber{5.65}&\overallnumber{20.23}&\overallnumber{7.06}&\overallnumber{11.85} &\overallnumber{3.36}&\overallnumber{6.40}&\overallnumber{21.66}&\overallnumber{7.73}&\overallnumber{12.86}\\
LTRGR \venueyear{(AAAI'24)} &\overallnumber{2.88}&\overallnumber{4.71}&\overallnumber{13.70}&\overallnumber{4.48}&\overallnumber{7.82} &\overallnumber{5.13}&\overallnumber{8.23}&\overallnumber{27.92}&\overallnumber{9.26}&\overallnumber{15.42} &\overallnumber{5.24}&\overallnumber{8.37}&\overallnumber{26.79}&\overallnumber{10.01}&\overallnumber{16.93}\\
RIPOR \venueyear{(WWW'24)} &\overallnumber{5.05}&\overallnumber{8.35}&\overallnumber{26.80}&\overallnumber{7.92}&\overallnumber{13.75} &\overallnumber{3.49}&\overallnumber{7.12}&\overallnumber{23.03}&\overallnumber{7.94}&\overallnumber{13.82} &\overallnumber{5.02}&\overallnumber{8.18}&\overallnumber{27.87}&\overallnumber{9.92}&\overallnumber{17.28}\\
MERGE \venueyear{(ACL'25)} &\overallnumber{5.68}&\overallnumber{9.26}&\overallnumber{29.74}&\overallnumber{9.05}&\overallnumber{15.17} &\overallnumber{5.80}&\overallnumber{9.25}&\overallnumber{30.86}&\overallnumber{10.20}&\overallnumber{17.45} &\overallnumber{5.21}&\overallnumber{8.90}&\overallnumber{27.64}&\overallnumber{10.95}&\overallnumber{17.60}\\
CaLIR \venueyear{(arXiv'26)} &\overallnumber[\underline]{7.50}&\overallnumber[\underline]{12.69}&\overallnumber[\underline]{37.40}&\overallnumber[\underline]{11.12}&\overallnumber[\underline]{18.79}&\overallnumber[\underline]{6.03}&\overallnumber[\underline]{10.48}&\overallnumber[\underline]{34.15}&\overallnumber[\underline]{12.43}&\overallnumber[\underline]{19.88}&\overallnumber[\underline]{6.05}&\overallnumber[\underline]{9.77}&\overallnumber[\underline]{30.89}&\overallnumber[\underline]{11.77}&\overallnumber[\underline]{18.42}\\
CAT-ID\textsuperscript{2} \venueyear{(WSDM'26)} &\overallnumber{6.11}&\overallnumber{8.89}&\overallnumber{29.03}&\overallnumber{8.96}&\overallnumber{14.53} &\overallnumber{5.70}&\overallnumber{9.34}&\overallnumber{31.44}&\overallnumber{10.68}&\overallnumber{18.01} &\overallnumber{5.36}&\overallnumber{8.87}&\overallnumber{28.82}&\overallnumber{10.34}&\overallnumber{17.14}\\
FORGE \venueyear{(KDD'26)} &\overallnumber{5.59}&\overallnumber{9.26}&\overallnumber{35.11}&\overallnumber{8.50}&\overallnumber{16.21} &\overallnumber{4.87}&\overallnumber{8.29}&\overallnumber{28.42}&\overallnumber{9.99}&\overallnumber{16.66} &\overallnumber{3.88}&\overallnumber{6.58}&\overallnumber{24.06}&\overallnumber{8.27}&\overallnumber{14.22}\\
HGRec \venueyear{(ICML'26)} &\overallnumber{6.70}&\overallnumber{11.41}&\overallnumber{34.61}&\overallnumber{10.19}&\overallnumber{17.32}
&\overallnumber{5.76}&\overallnumber{9.60}&\overallnumber{31.56}&\overallnumber{11.37}&\overallnumber{18.32}
&\overallnumber{3.78}&\overallnumber{6.18}&\overallnumber{24.37}&\overallnumber{8.40}&\overallnumber{13.83}\\

\midrule
\rowcolor{mainresultrow}\textbf{\method} &\overallnumber[\textbf]{9.28}&\overallnumber[\textbf]{14.34}&\overallnumber[\textbf]{39.58}&\overallnumber[\textbf]{12.99}&\overallnumber[\textbf]{20.79}&\overallnumber[\textbf]{6.86}&\overallnumber[\textbf]{11.76}&\overallnumber[\textbf]{35.38}&\overallnumber[\textbf]{14.22}&\overallnumber[\textbf]{21.54}&\overallnumber[\textbf]{7.61}&\overallnumber[\textbf]{11.63}&\overallnumber[\textbf]{32.86}&\overallnumber[\textbf]{14.18}&\overallnumber[\textbf]{20.79}\\
\rowcolor{mainimprovrow}\emph{Rel. Improv. (\%)} &\overallnumber{+23.73}&\overallnumber{+13.00}&\overallnumber{+5.83}&\overallnumber{+16.82}&\overallnumber{+10.64}&\overallnumber{+13.76}&\overallnumber{+12.21}&\overallnumber{+3.60}&\overallnumber{+14.40}&\overallnumber{+8.35}&\overallnumber{+25.79}&\overallnumber{+19.04}&\overallnumber{+6.38}&\overallnumber{+20.48}&\overallnumber{+12.87}\\
\bottomrule
  \end{tabular}}
\vspace{-1em}
\end{table}

%% file: tables/table_core_training.tex
\begin{table}[!t]
\centering\small
\newsavebox{\corestatbox}
\newcommand{\coremeanstd}[2]{\sbox{\corestatbox}{#1\textsubscript{\fontsize{6.5}{7}\selectfont\mdseries\textpm\,#2}}\makebox[\wd\corestatbox][r]{\raisebox{0pt}[\ht\corestatbox][\dp\corestatbox]{{\fontsize{9.35}{10}\selectfont #1\raisebox{-1.5pt}{\fontsize{6.75}{7}\selectfont\mdseries\textpm\kern0.3pt #2}}}}}
\setlength{\tabcolsep}{1.8pt}
\caption{Matched ESCI training objectives under a shared 1,200-update budget.
Entries are mean \textpm{} sample standard deviation across three paired seeds;
bold marks each column's highest mean overall.}
\label{tab:core-training}
\resizebox{\linewidth}{!}{\begin{tabular}{lrrrrrrrrrrrr}
\toprule
 & \multicolumn{4}{c}{ESCI-US} & \multicolumn{4}{c}{ESCI-ES} & \multicolumn{4}{c}{ESCI-JP}\\
\cmidrule(lr){2-5}\cmidrule(lr){6-9}\cmidrule(lr){10-13}
Training objective & R@10 & R@100 & N@10 & N@100
& R@10 & R@100 & N@10 & N@100
& R@10 & R@100 & N@10 & N@100\\
\midrule
Base-FT & \coremeanstd{12.82}{0.16} & \coremeanstd{37.60}{0.13} & \coremeanstd{11.25}{0.11} & \coremeanstd{18.92}{0.15} & \coremeanstd{10.55}{0.15} & \coremeanstd{34.22}{0.11} & \coremeanstd{12.53}{0.12} & \coremeanstd{19.95}{0.09} & \coremeanstd{9.85}{0.03} & \coremeanstd{30.98}{0.07} & \coremeanstd{11.90}{0.17} & \coremeanstd{18.55}{0.08}\\

RBD (\textit{H}=1) & \coremeanstd{13.23}{0.10} & \coremeanstd{38.20}{0.08} & \coremeanstd{11.78}{0.05} & \coremeanstd{19.44}{0.13} & \coremeanstd{10.84}{0.11} & \coremeanstd{34.46}{0.09} & \coremeanstd{12.98}{0.08} & \coremeanstd{20.30}{0.10} & \coremeanstd{10.29}{0.06} & \coremeanstd{31.48}{0.14} & \coremeanstd{12.50}{0.09} & \coremeanstd{19.10}{0.07}\\

Full-output KD & \coremeanstd{13.15}{0.07} & \coremeanstd{38.10}{0.13} & \coremeanstd{11.69}{0.10} & \coremeanstd{19.35}{0.09} & \coremeanstd{10.80}{0.05} & \coremeanstd{34.40}{0.11} & \coremeanstd{12.91}{0.06} & \coremeanstd{20.20}{0.14} & \coremeanstd{10.22}{0.09} & \coremeanstd{31.39}{0.08} & \coremeanstd{12.39}{0.11} & \coremeanstd{18.99}{0.12}\\

ResTD (\textit{H}=2) & \coremeanstd{13.77}{0.05} & \coremeanstd{38.83}{0.14} & \coremeanstd{12.32}{0.09} & \coremeanstd{20.08}{0.07} & \coremeanstd{11.28}{0.10} & \coremeanstd{34.91}{0.08} & \coremeanstd{13.55}{0.12} & \coremeanstd{20.86}{0.11} & \coremeanstd{10.95}{0.07} & \coremeanstd{32.04}{0.13} & \coremeanstd{13.24}{0.06} & \coremeanstd{19.83}{0.09}\\

RBD + future hard SID (\textit{H}=4) & \coremeanstd{13.56}{0.09} & \coremeanstd{38.68}{0.07} & \coremeanstd{12.18}{0.08} & \coremeanstd{19.93}{0.13} & \coremeanstd{11.16}{0.04} & \coremeanstd{34.81}{0.05} & \coremeanstd{13.40}{0.10} & \coremeanstd{20.71}{0.15} & \coremeanstd{10.81}{0.03} & \coremeanstd{31.96}{0.11} & \coremeanstd{13.12}{0.07} & \coremeanstd{19.72}{0.16}\\

Hard SID targets (\textit{H}=4) & \coremeanstd{13.10}{0.08} & \coremeanstd{38.12}{0.12} & \coremeanstd{11.70}{0.06} & \coremeanstd{19.30}{0.10} & \coremeanstd{10.78}{0.09} & \coremeanstd{34.40}{0.07} & \coremeanstd{12.84}{0.11} & \coremeanstd{20.19}{0.13} & \coremeanstd{10.20}{0.05} & \coremeanstd{31.39}{0.15} & \coremeanstd{12.35}{0.08} & \coremeanstd{18.98}{0.11}\\

Shuffled ResTD (\textit{H}=4) & \coremeanstd{12.86}{0.05} & \coremeanstd{37.68}{0.11} & \coremeanstd{11.22}{0.12} & \coremeanstd{18.95}{0.03} & \coremeanstd{10.51}{0.06} & \coremeanstd{34.10}{0.04} & \coremeanstd{12.60}{0.14} & \coremeanstd{19.92}{0.03} & \coremeanstd{9.89}{0.06} & \coremeanstd{31.02}{0.12} & \coremeanstd{11.88}{0.13} & \coremeanstd{18.59}{0.07}\\

Non-target output alignment & \coremeanstd{13.07}{0.06} & \coremeanstd{38.00}{0.11} & \coremeanstd{11.66}{0.12} & \coremeanstd{19.25}{0.07} & \coremeanstd{10.77}{0.10} & \coremeanstd{34.39}{0.14} & \coremeanstd{12.88}{0.09} & \coremeanstd{20.16}{0.12} & \coremeanstd{10.19}{0.08} & \coremeanstd{31.35}{0.13} & \coremeanstd{12.34}{0.06} & \coremeanstd{18.93}{0.10}\\

\midrule

\rowcolor{mainresultrow}\textbf{ResTD (\textit{H}=4)} & {\bfseries\coremeanstd{14.34}{0.04}} & {\bfseries\coremeanstd{39.58}{0.11}} & {\bfseries\coremeanstd{12.99}{0.07}} & {\bfseries\coremeanstd{20.79}{0.05}} & {\bfseries\coremeanstd{11.76}{0.04}} & {\bfseries\coremeanstd{35.38}{0.13}} & {\bfseries\coremeanstd{14.22}{0.07}} & {\bfseries\coremeanstd{21.54}{0.12}} & {\bfseries\coremeanstd{11.63}{0.17}} & {\bfseries\coremeanstd{32.86}{0.09}} & {\bfseries\coremeanstd{14.18}{0.08}} & {\bfseries\coremeanstd{20.79}{0.05}}\\
\bottomrule
\end{tabular}}
\end{table}

%% file: sections/teacher_geometry_analysis.tex
\phantomsection

\textbf{Dissecting Teacher Targets (RQ3).}
\label{sec:teacher-geometry-controls}Table~\ref{tab:teacher-geometry-controls} shows that our teacher attains the
highest mean score on every metric across all ESCI locales. We study controls
that preserve different combinations of assigned codes, distributional shape, codeword
preferences, and item--teacher correspondence at current and future levels.
Code-conditioned residual distributions outperform both future hard-SID
supervision and codebook-only targets, indicating that residual statistics
carry useful supervision beyond hard assignments and codebook distances.
Keeping the correspondence between codewords and their teacher probabilities
also improves retrieval over non-target permutation, which preserves the
probability profile but changes its association with codewords. Shape-matched
teachers drawn from other items still yield substantial gains over Base-FT,
showing that residual supervision captures structure shared across items.
ResTD improves further on every reported metric when each item retains its
own residual-induced preferences. Across these controls, residual geometry
provides useful supervision over competing codewords, while item-specific
preferences contribute an additional retrieval gain.
Appendix~\ref{app:teacher-geometry-controls} gives the control constructions,
matching quality, and retrieval results.
\input{tables/table_teacher_geometry_controls}

%% file: tables/table_teacher_geometry_controls.tex
\begin{table}[!htbp]
\centering
\fontsize{7}{8}\selectfont
\newlength{\teachershapereclaimed}
\newlength{\teachershapecompact}
\newlength{\teachershapeheading}
\settowidth{\teachershapereclaimed}{unmatched}
\settowidth{\teachershapecompact}{coarse}
\settowidth{\teachershapeheading}{Shape}
\ifdim\teachershapeheading>\teachershapecompact
  \setlength{\teachershapecompact}{\teachershapeheading}
\fi
\addtolength{\teachershapereclaimed}{-\teachershapecompact}
\setlength{\tabcolsep}{\dimexpr0.85pt+\teachershapereclaimed/34\relax}
\renewcommand{\arraystretch}{1.12}
\newsavebox{\teacherstatbox}
\newcommand{\teacherstat}[2]{\sbox{\teacherstatbox}{#1\textsubscript{\fontsize{4}{4}\selectfont\mdseries\textpm\,#2}}\makebox[\wd\teacherstatbox][r]{\raisebox{0pt}[\ht\teacherstatbox][\dp\teacherstatbox]{{\fontsize{7.2}{8}\selectfont #1\raisebox{-1pt}{\fontsize{4.4}{5}\selectfont\mdseries\textpm\kern0.1pt #2}}}}}
\newcommand{\teacherbeststat}[2]{{\bfseries\teacherstat{#1}{#2}}}

  \vspace{-1em}
  \caption{Retrieval performance of ResTD and teacher-target controls on ESCI (\%)
across three paired seeds. Values are mean~\textpm{}~SD; bold marks each column's
highest mean; \emph{unm.} denotes unmatched.}
\label{tab:teacher-geometry-controls}
\resizebox{\linewidth}{!}{\begin{tabular}{lccccrrrrrrrrrrrr}
\toprule
& & & & &
\multicolumn{4}{c}{ESCI-US}
& \multicolumn{4}{c}{ESCI-ES}
& \multicolumn{4}{c}{ESCI-JP}
\\

\cmidrule(lr){6-9}
\cmidrule(lr){10-13}
\cmidrule(lr){14-17}

Teacher objective
& Hard & Shape & Real & Pair
& R@10 & R@100 & N@10 & N@100
& R@10 & R@100 & N@10 & N@100
& R@10 & R@100 & N@10 & N@100
\\

\midrule

Base-FT
& $\checkmark$ & -- & -- & --
& \teacherstat{12.82}{0.16}
& \teacherstat{37.60}{0.13}
& \teacherstat{11.25}{0.11}
& \teacherstat{18.92}{0.15}
& \teacherstat{10.55}{0.15}
& \teacherstat{34.22}{0.11}
& \teacherstat{12.53}{0.12}
& \teacherstat{19.95}{0.09}
& \teacherstat{9.85}{0.03}
& \teacherstat{30.98}{0.07}
& \teacherstat{11.90}{0.17}
& \teacherstat{18.55}{0.08}
\\

Shuffled ResTD
& \texttimes{} & unm. & $\checkmark$ & \texttimes{}
& \teacherstat{12.86}{0.05}
& \teacherstat{37.68}{0.11}
& \teacherstat{11.22}{0.12}
& \teacherstat{18.95}{0.03}
& \teacherstat{10.51}{0.06}
& \teacherstat{34.10}{0.04}
& \teacherstat{12.60}{0.14}
& \teacherstat{19.92}{0.03}
& \teacherstat{9.89}{0.06}
& \teacherstat{31.02}{0.12}
& \teacherstat{11.88}{0.13}
& \teacherstat{18.59}{0.07}
\\

RBD + future hard SID
& $\checkmark$ & -- & -- & --
& \teacherstat{13.56}{0.09}
& \teacherstat{38.68}{0.07}
& \teacherstat{12.18}{0.08}
& \teacherstat{19.93}{0.13}
& \teacherstat{11.16}{0.04}
& \teacherstat{34.81}{0.05}
& \teacherstat{13.40}{0.10}
& \teacherstat{20.71}{0.15}
& \teacherstat{10.81}{0.03}
& \teacherstat{31.96}{0.11}
& \teacherstat{13.12}{0.07}
& \teacherstat{19.72}{0.16}
\\

Codebook-only (\textit{H}=4)
& $\checkmark$ & -- & \texttimes{} & \texttimes{}
& \teacherstat{13.48}{0.11}
& \teacherstat{38.55}{0.06}
& \teacherstat{12.05}{0.14}
& \teacherstat{19.81}{0.03}
& \teacherstat{10.98}{0.11}
& \teacherstat{34.65}{0.13}
& \teacherstat{13.22}{0.12}
& \teacherstat{20.51}{0.08}
& \teacherstat{10.63}{0.14}
& \teacherstat{31.76}{0.09}
& \teacherstat{12.95}{0.15}
& \teacherstat{19.51}{0.16}
\\

Code-conditioned mean
& $\checkmark$ & agg. & \texttimes{} & \texttimes{}
& \teacherstat{13.90}{0.08}
& \teacherstat{39.03}{0.15}
& \teacherstat{12.50}{0.10}
& \teacherstat{20.27}{0.16}
& \teacherstat{11.39}{0.06}
& \teacherstat{35.00}{0.08}
& \teacherstat{13.66}{0.03}
& \teacherstat{20.97}{0.13}
& \teacherstat{11.09}{0.13}
& \teacherstat{32.24}{0.10}
& \teacherstat{13.47}{0.05}
& \teacherstat{20.05}{0.15}
\\

Same-code + entropy-bin shuffle
& $\checkmark$ & coarse & $\checkmark$ & \texttimes{}
& \teacherstat{13.82}{0.07}
& \teacherstat{38.96}{0.10}
& \teacherstat{12.44}{0.06}
& \teacherstat{20.22}{0.11}
& \teacherstat{11.29}{0.11}
& \teacherstat{34.89}{0.07}
& \teacherstat{13.60}{0.09}
& \teacherstat{20.91}{0.12}
& \teacherstat{11.03}{0.04}
& \teacherstat{32.17}{0.13}
& \teacherstat{13.44}{0.10}
& \teacherstat{20.03}{0.05}
\\

Non-target permutation
& $\checkmark$ & exact & \texttimes{} & \texttimes{}
& \teacherstat{13.76}{0.10}
& \teacherstat{38.86}{0.16}
& \teacherstat{12.36}{0.15}
& \teacherstat{20.12}{0.09}
& \teacherstat{11.29}{0.04}
& \teacherstat{34.93}{0.15}
& \teacherstat{13.54}{0.12}
& \teacherstat{20.84}{0.06}
& \teacherstat{10.96}{0.04}
& \teacherstat{32.11}{0.05}
& \teacherstat{13.31}{0.10}
& \teacherstat{19.89}{0.12}
\\

Shape-matched teacher swap
& $\checkmark$ & near & $\checkmark$ & \texttimes{}
& \teacherstat{14.16}{0.10}
& \teacherstat{39.36}{0.07}
& \teacherstat{12.79}{0.03}
& \teacherstat{20.58}{0.11}
& \teacherstat{11.61}{0.11}
& \teacherstat{35.23}{0.03}
& \teacherstat{14.00}{0.05}
& \teacherstat{21.31}{0.07}
& \teacherstat{11.41}{0.04}
& \teacherstat{32.61}{0.07}
& \teacherstat{13.90}{0.09}
& \teacherstat{20.49}{0.17}
\\

\midrule

\rowcolor{mainresultrow}
\textbf{ResTD (\textit{H}=4)}
& $\checkmark$
& exact
& $\checkmark$
& $\checkmark$
& \teacherbeststat{14.34}{0.04}
& \teacherbeststat{39.58}{0.11}
& \teacherbeststat{12.99}{0.07}
& \teacherbeststat{20.79}{0.05}
& \teacherbeststat{11.76}{0.04}
& \teacherbeststat{35.38}{0.13}
& \teacherbeststat{14.22}{0.07}
& \teacherbeststat{21.54}{0.12}
& \teacherbeststat{11.63}{0.17}
& \teacherbeststat{32.86}{0.09}
& \teacherbeststat{14.18}{0.08}
& \teacherbeststat{20.79}{0.05}
\\

\bottomrule

\end{tabular}}
\vspace{-1em}

\end{table}

%% file: sections/future_probe_analysis.tex
We fit matched-capacity probes to frozen retrievers to recover
collision-corrected teacher distributions from decoder states. Across
conditions, query splits, probe architecture, initialization, and fitting
procedure are fixed for the three retriever seeds, with a separate probe for
each target offset.
Figure~\ref{fig:future-probe} shows that ResTD ($H=4$) reduces KL relative to
current-level RBD ($H=1$) at every future state--target pair across all three
locales, with relative reductions of 5.3--13.8\%. In contrast, reductions on
current targets remain small, at 0.1--0.3\%. The largest reduction in each
locale occurs for the longest-horizon pair, where the first decoder state is
probed against the fourth-level teacher distribution. This pattern localizes
the effect of multi-horizon supervision to future codebook information:
later-level preferences become substantially more recoverable from earlier
decoder states, while current-target probe performance remains stable.
Matched retrieval results are reported in
Section~\ref{sec:training-comparisons}, and probe specifications are provided
in Appendix~\ref{app:future-probe}.

%% file: sections/multipositive_analysis.tex
\phantomsection
\textbf{Retrieval Gains across Query Groups (RQ5).}
\label{sec:multipositive}
We study how the gains from ResTD vary with disagreement among the teachers
associated with relevant items. Figure~\ref{fig:disagreement-gains} partitions
multi-positive queries by first-level teacher JS divergence and reports the
corresponding changes in Recall@100 and NDCG@10. On ESCI-US and ESCI-JP,
improvements over both Base-FT and RBD ($H=1$) increase with teacher
disagreement, with the largest gains appearing in the high-disagreement group.
On ESCI-ES, NDCG@10 gains peak in the intermediate-disagreement group.
Across locales, stratified trends remain metric dependent, indicating no
uniform gain pattern from disagreement. Controlling for the number of relevant
items and initial query difficulty preserves positive associations on US and
JP; the ES confidence intervals span zero. These query-level comparisons
characterize dataset- and metric-specific associations between teacher
disagreement and retrieval gains. Detailed teacher-disagreement statistics,
disagreement-stratified gains, and adjusted analyses are reported in
Appendix~\ref{app:multipositive-details}.

%% file: sections/generative_recommendation.tex
\begingroup
\setlength{\columnsep}{12pt}
\setlength{\intextsep}{4pt}
\begin{wraptable}{r}{0.49\textwidth}
\vspace{-10pt}
\centering
\caption{Generative recommendation results (\%). ResTD uses TIGER as its
backbone; improvement is relative to the best baseline per column.}
\label{tab:generative-recommendation}
\scriptsize
\setlength{\tabcolsep}{1.5pt}
\begin{tabular*}{\linewidth}{@{\extracolsep{\fill}}lrrrrrr@{}}
\toprule
\multirow{2}{*}{\textbf{Model}} & \multicolumn{2}{c}{\textbf{B-Shop}} & \multicolumn{2}{c}{\textbf{I-Shop}} & \multicolumn{2}{c}{\textbf{Yelp}} \\
\cmidrule(lr){2-3}\cmidrule(lr){4-5}\cmidrule(lr){6-7}
& \textbf{R@10} & \textbf{N@10} & \textbf{R@10} & \textbf{N@10} & \textbf{R@10} & \textbf{N@10} \\
\midrule
SASRec & 5.88 & 3.13 & 9.47 & 6.90 & 2.96 & 1.52 \\
BigRec & 2.99 & 1.98 & 5.76 & 4.91 & 1.69 & 1.42 \\
P5-SID & 5.84 & 3.35 & 9.64 & 7.30 & 3.24 & 1.70 \\
P5-CID & 5.97 & 3.47 & 9.87 & 7.51 & 3.47 & 1.81 \\
LETTER & \underline{6.72} & \underline{3.64} & \underline{11.22} & \underline{8.31} & \underline{4.26} & \underline{2.31} \\
TIGER & 6.10 & 3.31 & 10.58 & 7.97 & 4.07 & 2.13 \\
ETEGRec & 6.15 & 3.35 & 11.06 & 8.10 & 4.15 & 2.14 \\
\midrule
\rowcolor{mainresultrow}\textbf{\method} & \textbf{7.28} & \textbf{4.00} & \textbf{11.42} & \textbf{8.45} & \textbf{4.87} & \textbf{2.66} \\
\rowcolor{mainimprovrow}\emph{Improv. (\%)} & +8.33 & +9.89 & +1.78 & +1.68 & +14.32 & +15.15 \\
\bottomrule
\end{tabular*}
\vspace{-4pt}
\end{wraptable}

\Needspace{9\baselineskip}
\phantomsection
\textbf{Generative Recommendation (RQ6).}
\label{sec:generative-recommendation}
We apply ResTD to TIGER for history-conditioned next-item recommendation,
using the frozen RQ tokenizer to supervise decoder states predicting semantic
item codes. Table~\ref{tab:generative-recommendation} reports Recall@10 and
NDCG@10 on Beauty (B-Shop), Instruments (I-Shop), and Yelp. ResTD achieves the
best performance across all dataset--metric settings, improving over strong
baselines on e-commerce and non-e-commerce benchmarks. Gains are consistently strongest on B-Shop and Yelp, and positive across both I-Shop metrics.
The results demonstrate that residual trajectory supervision extends beyond
ESCI retrieval and provides effective signals for TIGER-based generative
recommendation. Component effects are examined through retrieval experiments.
Implementation details are provided in
Appendix~\ref{app:recommendation-setup}.

\par
\ifnum\value{WF@wrappedlines}>1
  \vspace{\dimexpr\value{WF@wrappedlines}\baselineskip-\baselineskip\relax}
\fi
\WFclear
\endgroup

%% file: sections/case_study_analysis.tex
\begin{figure}[!htbp]
  \centering
  \includegraphics[width=\linewidth]{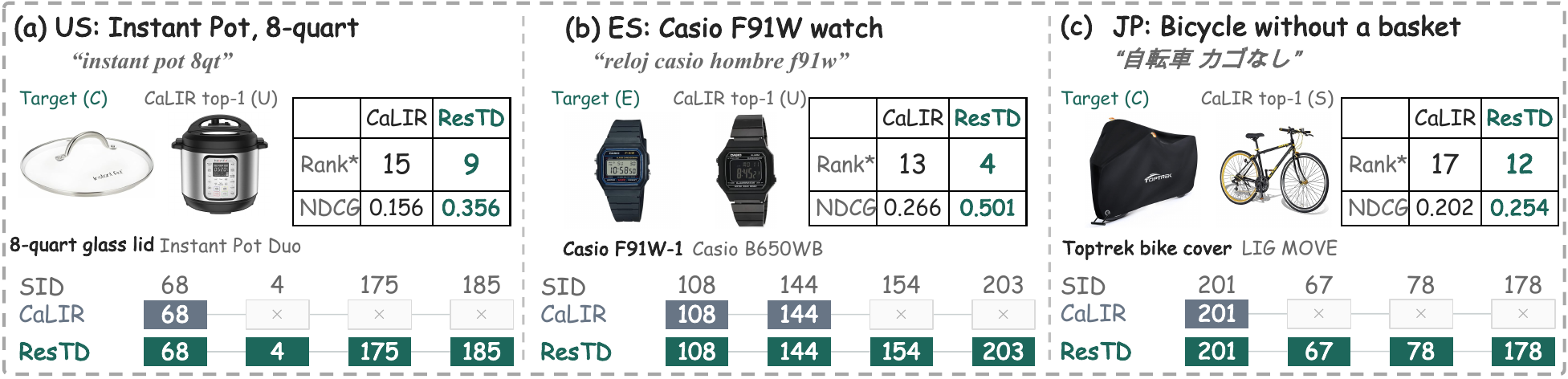}
\caption{Retrieval examples on ESCI. Rank* is the exhaustive SID rank;
NDCG@10 and SID traces use beam width 10. E/S/C/U:
Exact/Substitute/Complement/unjudged; $\times$ marks pruned prefixes.}
  \label{fig:rq-case-study}
  \vspace{-1em}
\end{figure}

\phantomsection

\textbf{Qualitative Analysis (RQ7).}
\label{sec:qualitative-analysis}Figure~\ref{fig:rq-case-study} illustrates how ResTD improves retrieval by
preserving promising identifier paths during generation. In the Casio example,
the baseline prunes the target identifier early, whereas ResTD retains the
complete path and improves the ranking of the target item. Similar behavior is
observed for the complementary products in the US and JP examples, where ResTD
maintains relevant candidates through later decoding steps and improves ranking
quality.
These examples demonstrate the benefit of residual trajectory supervision.
Future quantization preferences provide additional guidance for early
identifier generation, allowing relevant items to remain reachable during
autoregressive retrieval. Additional examples are provided in
Appendix~\ref{app:case-study}.

%% file: sections/appendix_layout.tex
\newcommand{\appendixtablestyle}{\fontsize{9}{10.5}\selectfont
  \setlength{\tabcolsep}{4pt}\renewcommand{\arraystretch}{1.12}}
\newcommand{\appendixstat}[2]{\mbox{#1\textsubscript{\fontsize{6.5}{7}\selectfont\mdseries\textpm\,#2}}}
\newcommand{\appendixbeststat}[2]{\mbox{\textbf{#1}\textsubscript{\fontsize{6.5}{7}\selectfont\mdseries\textpm\,#2}}}
\setlength{\textfloatsep}{12pt plus 2pt minus 2pt}
\setlength{\floatsep}{10pt plus 2pt minus 2pt}
\setlength{\intextsep}{10pt plus 2pt minus 2pt}
\clubpenalty=10000
\widowpenalty=10000
\displaywidowpenalty=10000
\renewcommand{\topfraction}{0.95}
\renewcommand{\bottomfraction}{0.90}
\renewcommand{\textfraction}{0.05}
\renewcommand{\floatpagefraction}{0.75}
\setcounter{topnumber}{3}
\setcounter{bottomnumber}{2}
\setcounter{totalnumber}{5}
\setcounter{tocdepth}{2}
\makeatletter
\setlength{\@fptop}{0pt}
\setlength{\@fpsep}{12pt plus 2pt minus 2pt}
\setlength{\@fpbot}{0pt plus 1fil}
\renewcommand{\l@section}[2]{\addpenalty{\@secpenalty}\addvspace{5pt}{\bfseries\@dottedtocline{1}{0em}{1.7em}{#1}{#2}}}
\renewcommand{\l@subsection}{\@dottedtocline{2}{1.7em}{2.5em}}
\makeatother

%% file: sections/appendix_experimental_setup.tex
\section{Experimental Setup and Implementation}
\label{app:experimental-setup}
The main retrieval experiments use the multilingual ESCI benchmark.
Appendix~\ref{app:nq320k} reports additional document retrieval experiments
on NQ320K.
Appendix~\ref{app:recommendation-setup} describes the TIGER-based
recommendation experiments for RQ6.

\subsection{Retrieval Datasets}
ESCI provides queries and product-level relevance judgments in multiple
languages \citep{reddy2022esci}. We follow the filtering and data splits of
\citet{zhang2026calir}, obtaining 288,372 products for ESCI-US, 101,957 for
ESCI-ES, and 119,052 for ESCI-JP. After filtering and query deduplication,
the test sets contain 6,014, 1,656, and 1,883 queries, respectively.
Each query may have several relevant products. Evaluation follows the
same relevance definitions and reports Recall at cutoffs 5, 10, and 100
and NDCG at cutoffs 10 and 100.

\subsection{Retrieval Implementation Details}
\paragraph{Model and index.}
We use T5-base for ESCI-US and mT5-base for ESCI-ES and ESCI-JP.
The collision-resolved SID index comprises four quantization levels with
256 codewords per level. Residual targets are constructed along the final
SID paths, as described in Section~\ref{sec:collision}. All controlled
comparisons train the retriever with a fixed index and evaluation protocol.

\paragraph{Default configuration.}
All three ESCI locales use a common configuration: $H=4$,
$\lambda_{\max}=0.1$, $\tau_T=\tau_S=0.2$, $\rho=0.7$, and
collision-correction floor $\epsilon=0.1$ with margin $\delta=0.001$.
Each fine-tuning condition includes four auxiliary heads, so varying $H$ changes the
supervised horizons while preserving the architecture.
Appendix~\ref{app:hyperparameter-sensitivity} examines departures from
these defaults. Appendix~\ref{app:collision-correction-sensitivity}
evaluates the correction rule and its mixing-weight floor.

\paragraph{Optimization.}
Each condition is trained for 1,200 optimizer updates with an effective
batch size of 128, microbatches of 8 examples, a learning rate of $10^{-5}$,
and cosine decay. The learning rate and auxiliary coefficient warm up over
the first 120 updates. A multiplicative factor, treated as constant during
backpropagation, limits the weighted auxiliary loss to 5\% of the SID loss.
We use seeds 42, 2027, and 2028. Within each seed, conditions share
initial retriever parameters, training data, minibatch order, SID index,
frozen category adapter, optimization schedule, and evaluation protocol.

\subsection{Evaluation Protocol and Uncertainty}
\label{app:evaluation-protocol}
\paragraph{Decoding.}
All fine-tuning comparisons use trie-constrained decoding with beam width
100 and category width three. The category-prediction adapter is frozen
and shared across objectives. All reported means and confidence intervals
use this decoding protocol.

\paragraph{SID-level relevance and ranking.}
\label{app:sid-evaluation}
Our CaLIR reproduction and all subsequent fine-tuning conditions are
evaluated at the SID level. For query $q$, products labeled Exact,
Substitute, or Complement receive binary relevance
$\mathrm{rel}(q,d)=1$; Irrelevant and unjudged items receive
$\mathrm{rel}(q,d)=0$. Let $y(d)$ be the complete SID of item $d$. For items
sharing a SID, we take the maximum relevance:
\begin{equation}
\mathrm{rel}(q,y)=\max_{d:y(d)=y}\mathrm{rel}(q,d),
\qquad
\mathcal Y^+(q)=\{y(d):d\text{ is positive for }q\}.
\label{eq:sid-relevance}
\end{equation}
Predictions are sorted by retrieval score, retaining only the
highest-ranked occurrence of each complete SID. For the top $k$ distinct
SIDs, denoted by $\widehat{\mathcal Y}_{1:k}(q)$, Recall is
\begin{equation}
\mathrm{Recall}@k(q)
=100\frac{|\widehat{\mathcal Y}_{1:k}(q)\cap\mathcal Y^+(q)|}
{|\mathcal Y^+(q)|}.
\label{eq:sid-recall}
\end{equation}
NDCG uses the same unique-SID ranking and binary relevance judgments.
Let $\hat y_j$ be the SID at rank $j$ in the deduplicated list. The
discounted cumulative gain and its ideal value are
\begin{equation}
\mathrm{DCG}@k(q)
=\sum_{j=1}^{k}\frac{\mathrm{rel}(q,\hat y_j)}{\log_2(j+1)},
\qquad
\mathrm{IDCG}@k(q)
=\sum_{j=1}^{\min(k,|\mathcal Y^+(q)|)}\frac{1}{\log_2(j+1)}.
\label{eq:sid-dcg}
\end{equation}
The ideal ranking places all unique positive SIDs first. Unjudged SIDs
and unfilled ranks contribute zero to DCG. The normalized score is
\begin{equation}
\mathrm{NDCG}@k(q)
=100\frac{\mathrm{DCG}@k(q)}{\mathrm{IDCG}@k(q)}.
\label{eq:sid-ndcg}
\end{equation}
Both Recall and NDCG are averaged over test queries.

\paragraph{Metric units.}
All reported Recall and NDCG values use the $[0,100]$ percentage scale,
except the query-level NDCG@10 values in Figure~\ref{fig:rq-case-study}
and NDCG@100 values in Figure~\ref{fig:case-study}, which use $[0,1]$.
Let $M(q;\theta)$ denote a per-query ESCI score on the percentage scale.
A difference between model scores is therefore measured in percentage points.
The relative improvements in Table~\ref{tab:main} are computed as
$100(M_{\mathrm{ResTD}}-M_{\mathrm{base}})/M_{\mathrm{base}}$ from the displayed
scores, using the strongest baseline in each column and rounding the result
to two decimal places. The reference is the reproduced CaLIR in all columns.
Appendix~\ref{app:multipositive-details} analyzes per-query gains in
percentage points through stratification and regression.

\paragraph{Controlled comparisons.}
Full-output KD, non-target output alignment, and horizon comparisons use
the same three paired seeds and experimental controls.
Appendix~\ref{app:full-output-kd} defines the full-output objective and its
comparison with current-level distillation. Summaries include all three seeds.

\paragraph{Evaluation and uncertainty.}
Table~\ref{tab:core-training} reports means and standard deviations across
three training seeds. The ResTD ($H=4$) results in Table~\ref{tab:main}
are calculated from these same runs. Paired differences quantify variation
across training seeds, whereas query resampling characterizes uncertainty
conditional on the fitted models. Matched future-target controls assess
multi-horizon supervision.

\subsection{Baselines and Matched Controls}
\label{app:baselines}
Table~\ref{tab:main} compares ResTD with sparse, dense, and generative
retrieval methods. Sparse retrieval relies on lexical matching, dense
retrieval compares continuous representations, and generative retrieval
predicts document or item identifiers. External baseline scores follow
their published protocols. The CaLIR scores come from our reproduction.
ResTD uses the three-seed $H=4$ means in Table~\ref{tab:core-training}.

\subsubsection{Sparse retrieval}
\begin{itemize}
\setlength{\itemsep}{2pt}
\setlength{\parsep}{0pt}
    \item \textbf{BM25} is a probabilistic lexical model that scores query--item pairs using term frequency and inverse document frequency, with term-frequency saturation and document-length normalization. It requires no learned representations~\citep{robertson2009probabilistic}.
\end{itemize}
\subsubsection{Dense retrieval}
\begin{itemize}
\setlength{\itemsep}{2pt}
\setlength{\parsep}{0pt}
    \item \textbf{DPR} encodes queries and items separately in a shared vector space and ranks candidates by representation similarity. Contrastive training uses positive query--item pairs and hard or in-batch negatives~\citep{karpukhin2020dense}.
    \item \textbf{MPNet} combines masked and permuted language modeling during pretraining, learning from bidirectional context while modeling dependencies among predicted tokens. Its sentence representations support dense retrieval~\citep{song2020mpnet}.
    \item \textbf{Sentence-T5} learns sentence embeddings by pooling T5 encoder states and training semantically related texts to have similar representations. The resulting embeddings support dense retrieval based on pretrained sequence-to-sequence representations~\citep{ni2022sentencet5}.
    \item \textbf{Sentence-mT5} extends Sentence-T5 to multilingual mT5 encoders and multilingual sentence-embedding data. We use it as the corresponding dense baseline for the Spanish and Japanese ESCI datasets~\citep{yano-etal-2024-multilingual}.
    \item \textbf{BGE-M3} supports multilingual dense, sparse, and multi-vector retrieval at different input granularities. Its self-knowledge-distillation objective combines supervision from these retrieval functions. Our comparison uses its dense representations~\citep{chen2024m3}.
    \item \textbf{LaSER} distills explicit reasoning into dense retrieval representations. Dual-view self-distillation aligns explicit and latent reasoning through output and intermediate-trajectory objectives, removing the need to generate a rationale at inference~\citep{jin2026laser}.
\end{itemize}
\subsubsection{Generative retrieval}
\begin{itemize}
\setlength{\itemsep}{2pt}
\setlength{\parsep}{0pt}
    \item \textbf{DSI\textsubscript{naive}} assigns arbitrary numerical document identifiers as tokenized strings. A text-to-text Transformer maps queries to these identifiers, encoding the corpus in its parameters without imposing semantics on the identifiers~\citep{tay2022transformer}.
    \item \textbf{DSI\textsubscript{semantic}} generates document identifiers constructed from semantic representations. Similar documents can share parts of their identifiers, introducing corpus structure into the output space while preserving a discrete address for each document~\citep{tay2022transformer}.
    \item \textbf{TIGER} represents each item as a sequence of semantic codewords obtained by residual quantization. A Transformer predicts the next item's identifier autoregressively from the user's interaction history~\citep{rajput2023recommender}.
    \item \textbf{Hi-Gen} learns representations from semantic relevance and e-commerce efficiency signals, then builds identifiers with category-guided hierarchical clustering. Its position-aware loss models unequal importance and dependencies across identifier positions~\citep{wu2024higen}.
    \item \textbf{LTRGR} addresses the mismatch between token-level generation and document ranking. An additional learning-to-rank stage optimizes the generator for retrieval quality without changing the inference procedure~\citep{li2024ltrgr}.
    \item \textbf{RIPOR} constructs identifiers by quantizing relevance-aware document representations and trains them with a prefix-oriented ranking objective. This objective improves the ranking of intermediate prefixes during autoregressive beam search~\citep{zeng2024ripor}.
    \item \textbf{CAT-ID$^2$} integrates an e-commerce category tree into semantic-identifier learning. Its hierarchical class-constraint, cluster-scale-constraint, and dispersion losses encourage category-consistent codes, balanced codeword usage, and separation of reconstructed items while preserving identifier uniqueness~\citep{liu2026catid}.
    \item \textbf{MERGE} uses query--document relevance to supervise document-identifier learning. A multi-relevance alignment module, an outer contrastive objective for binary relevance, and an inner objective for graded relevance jointly encode hierarchical relevance information while maintaining distinct identifiers~\citep{zhang2025merge}.
    \item \textbf{FORGE} studies semantic-identifier construction for industrial generative retrieval. It categorizes design choices and proposes inexpensive quality measures correlated with downstream retrieval performance. We use its reported configuration~\citep{fu2026forge}.
    \item \textbf{HGRec} combines a hyperbolic RQ-VAE with differential-length codebooks for generative recommendation. Hyperbolic representations model the hierarchy across quantization levels, and a pyramidal codebook-size schedule allocates capacity accordingly, reducing collisions and improving codebook utilization~\citep{zhang2026hgrec}.
    \item \textbf{CaLIR} (Category-guided Latent Intent Reasoning) adds category-guided latent intent states before SID decoding. It links shopping intent to SIDs via hierarchical semantic reasoning, query-wise reasoning enhancement for multi-positive queries, and a query-specific dynamic prefix trie with reasoning-aware constrained decoding~\citep{zhang2026calir}.
\end{itemize}

\subsubsection{Matched training controls}
Within each locale, all controlled comparisons start from our reproduced
CaLIR model. Base-FT optimizes the original retrieval objective, and
RBD ($H=1$) adds current-level residual supervision. Multi-horizon variants
also supervise future levels with residual distributions or hard SID
targets. Shuffled ResTD removes the original item--teacher pairing.
Full-output KD applies the current-level teacher to the retriever's SID
output distribution, while non-target output alignment supervises the
relative probabilities of alternative codes. Section~\ref{sec:training-comparisons} and
Appendix~\ref{app:output-supervision} give the corresponding controlled
comparisons and objective definitions.

\input{sections/appendix_recommendation_setup}

%% file: sections/appendix_recommendation_setup.tex
\subsection{Generative Recommendation Setup}
\label{app:recommendation-setup}

\paragraph{Datasets and split.}
We evaluate recommendation on Beauty (B-Shop), Instruments (I-Shop), and
Yelp after 5-core filtering, retaining users and items with at least five
interactions. Dataset statistics are given in
Table~\ref{tab:recommendation-data}. We order each user's interactions
chronologically. The penultimate interaction is held out for validation,
and the final interaction for testing. Earlier interactions form the training
set. Each example
pairs a history prefix with the next item, and each history contains only
interactions preceding its target. Validation and test targets are excluded
from training.

\begin{table}[!htbp]
\centering
\caption{Recommendation datasets after preprocessing. Average length is the
number of interactions per user, measured before the leave-one-out split.}
\label{tab:recommendation-data}
\appendixtablestyle
\begin{tabular*}{\linewidth}{@{\extracolsep{\fill}}lrrrr@{}}
\toprule
\textbf{Dataset} & \textbf{Users} & \textbf{Items} &
\textbf{Interactions} & \textbf{Avg. length} \\
\midrule
Beauty (B-Shop) & 22,363 & 12,101 & 198,502 & 8.88 \\
Instruments (I-Shop) & 24,772 & 9,922 & 206,153 & 8.32 \\
Yelp & 30,431 & 20,033 & 304,524 & 10.01 \\
\bottomrule
\end{tabular*}
\end{table}

\paragraph{Item content and SID construction.}
For Beauty and Instruments, LLaMA-7B generates item content from product
titles and descriptions. Yelp item representations use business names
and categories; review text supplies the interaction records.
A pretrained RQ-VAE maps content
representations to three semantic codes, with 256 codewords per level.
A disambiguation token distinguishes items with the same three-code
sequence, producing the four-token SID
$\vy(d)=(y_1,y_2,y_3,y_{\mathrm{dis}})$.
The tokenizer, codebooks, and item-to-SID mapping remain frozen throughout
recommender training.

\paragraph{TIGER backbone and recommendation objective.}
The recommendation model follows TIGER: a T5-style encoder--decoder with
six layers in each component and a hidden size of 128. The encoder
represents the chronological SID sequence of previously interacted items,
and the decoder predicts the next item's SID autoregressively. For an
interaction history $\mathcal H_u$ and target item $d^+$, we minimize the
negative log-likelihood
\begin{equation}
\mathcal L_{\mathrm{rec}}
=-\sum_{t=1}^{4}\log p_\theta
\bigl(y_t(d^+)\mid\mathcal H_u,y_{<t}(d^+)\bigr),
\qquad y_4=y_{\mathrm{dis}}.
\label{eq:recommendation-loss}
\end{equation}
This history-conditioned loss replaces the retrieval objective in
Equation~\ref{eq:total}.

\paragraph{Residual trajectory supervision.}
Under teacher forcing, $\vh_t$ is the final decoder-layer state that predicts
the target item's $t$-th semantic code. The frozen tokenizer supplies the
target item's pre-assignment residuals and codebooks for the teacher in
Equation~\ref{eq:mhteacher}; horizon-specific projections of $\vh_t$ define
the student in Equation~\ref{eq:mhstudent}. The student conditions on
$\mathcal H_u$, and all teacher targets come from $d^+$.
We apply Equation~\ref{eq:retrace} only to the three RQ levels: valid
state--horizon pairs satisfy $1\leq t\leq3$ and $0\leq s\leq3-t$,
corresponding to $L=H=3$.
The first semantic-code state is thus supervised by all three quantization
levels, whereas the last is supervised only by its current-level target.
The disambiguation token has no associated residual codebook. It receives
only the original generation loss and is excluded from the trajectory
loss and its normalizer.
The training objective is
$\mathcal L_{\mathrm{rec}}+\lambda(u)\mathcal L_{\mathrm{RT}}$;
only the recommender and auxiliary projections are optimized. The item
tokenizer remains frozen. Inference follows TIGER with the fixed SID
index and no auxiliary heads.

\paragraph{Auxiliary configuration.}
We use the shared ResTD settings: $\lambda_{\max}=0.1$,
$\tau_T=\tau_S=0.2$, and $\rho=0.7$. The auxiliary coefficient follows
the linear warmup in Equation~\ref{eq:total}. A multiplicative factor,
held constant during backpropagation, caps the weighted trajectory loss at 5\% of
$\mathcal L_{\mathrm{rec}}$. Teacher construction follows
Section~\ref{sec:collision} at the three RQ levels, with
floor $\epsilon=0.1$ and margin $\delta=0.001$.

\paragraph{Optimization and evaluation.}
We use AdamW with weight decay 0.05 and select the recommender learning rate
on the validation set from $\{0.01,0.005,0.001,0.0005\}$.
Validation performance also determines early stopping. This optimization
protocol is specific to recommendation and differs from the fixed-budget
ESCI fine-tuning protocol. At evaluation, the held-out next item is ranked
against the full catalog. We report test Recall@10 and NDCG@10 on the
$[0,100]$ scale.
For a single held-out item at rank $r$, the per-user scores are
$100\,\mathbf{1}[r\leq10]$ and
$100\,\mathbf{1}[r\leq10]/\log_2(r+1)$, respectively; the reported metrics
average these scores over users.

\paragraph{Comparison scope.}
Table~\ref{tab:generative-recommendation} compares ResTD on the TIGER
architecture with the TIGER baseline and other recommendation methods.
These experiments assess ResTD's applicability to recommendation.
Controlled ESCI comparisons examine the contributions of its individual
components.

%% file: sections/appendix_geometry_details.tex
\subsection{Information Retained by Residual Teachers}
\label{app:geometry-details}

A hard assignment records the nearest codeword. Residual distances to
competing codewords can vary within the same quantization cell.

\begin{proposition}
\label{prop:lossy}
Let $\mathcal{C}=\{\vc_j\}_{j=1}^{K}$ contain at least two distinct
codewords. For any finite temperature $\tau>0$, there exist
$\vr\neq\vr'$ such that
\[
\begin{gathered}
\argmin_j \|\vr-\vc_j\|_2^2
  =
  \argmin_j \|\vr'-\vc_j\|_2^2,
\\[3pt]
 \gQ_\tau(\cdot\mid\vr)
  \neq
  \gQ_\tau(\cdot\mid\vr'),
  \qquad
  \gQ_\tau(j\mid\vr)
  \propto
  \exp\!\left(
    -\frac{\|\vr-\vc_j\|_2^2}{\tau}
  \right).
\end{gathered}
\]
\end{proposition}

\begin{proof}
Choose $a,b$ with $\vc_a\neq\vc_b$, and set $\vr=\vc_a$ and
$\vr'=\vc_a+\varepsilon(\vc_b-\vc_a)$. For sufficiently small
$\varepsilon>0$, both residuals have the same set of nearest-code indices,
namely those with $\vc_j=\vc_a$. Their $b$-versus-$a$ log-odds differ by
$2\varepsilon\|\vc_b-\vc_a\|_2^2/\tau>0$, so their teacher distributions
cannot be equal.
\end{proof}

Within a Voronoi cell, the winning code is fixed while relative codeword
distances vary \citep{gray1984vector}. The residual $\vr_{t-1}$ induces
a distribution $\gQ_t(j\mid d)$ at temperature $\tau_T$, with pairwise log-odds
\begin{equation}
  \log
  \frac{\gQ_t(j\mid d)}
       {\gQ_t(k\mid d)}
  =
  \frac{
    2\vr_{t-1}^{\top}(\vc_{t,j}-\vc_{t,k})
    +\|\vc_{t,k}\|_2^2
    -\|\vc_{t,j}\|_2^2
  }{\tau_T}.
  \label{eq:logodds}
\end{equation}
These log-odds preserve the ordering and margins among competing codes,
information absent from hard labels and their uniformly smoothed
counterparts \citep{szegedy2016rethinking}. Residual trajectories therefore
provide item-specific supervision in the indexer's codebook space.
\par

\subsection{Gradients of Hard and Residual Supervision}
\label{app:supervision-gradients}

For a valid state--horizon pair $(t,s)$ at target level
$\ell=t+s$, let $\vu=g_s(\vh_t)$, $a_j=-\|\vu-\vc_{\ell,j}\|_2^2/\tau_S$,
and $\gP=\softmax(a)$. The corrected teacher $\widetilde{\gQ}$ is fixed with
respect to the student. For the assigned code $y=y_\ell^{\mathrm{final}}$,
define $\mathcal L_{\mathrm{soft}}=\KL(\widetilde{\gQ}\|\gP)$ and
$\mathcal L_{\mathrm{hard}}=-\log \gP_y$, with logit gradients
\begin{equation}
  \frac{\partial\mathcal L_{\mathrm{soft}}}{\partial a_j}
  =\gP_j-\widetilde{\gQ}_j,
  \qquad
  \frac{\partial\mathcal L_{\mathrm{hard}}}{\partial a_j}
  =\gP_j-\mathbf{1}[j=y].
  \label{eq:target-gradients}
\end{equation}
Because teacher entropy is constant during optimization, KL divergence
and soft-target cross-entropy induce identical student gradients. At a
fixed prediction, replacing the one-hot target changes the gradient by
$\mathbf e_y-\widetilde{\gQ}$. The change depends on how the teacher assigns
probability to individual codewords, which is not determined by the hard
code or teacher entropy alone.

For the distance-based head, the chain rule gives
\begin{align}
  \nabla_{\vu}\mathcal L_{\mathrm{soft}}
  &=\sum_j(\gP_j-\widetilde{\gQ}_j)
    \frac{-2(\vu-\vc_{\ell,j})}{\tau_S}
    \nonumber\\
  &=\frac{2}{\tau_S}
    \left(\sum_j \gP_j\vc_{\ell,j}
    -\sum_j\widetilde{\gQ}_j\vc_{\ell,j}\right),
  \label{eq:codebook-gradient}
\end{align}
where $\sum_j(\gP_j-\widetilde{\gQ}_j)=0$. For a hard target, the teacher
expectation reduces to $\vc_{\ell,y}$. The difference between student and
teacher codebook expectations is propagated through the Jacobian of $g_s$
to the shared decoder state $\vh_t$, with weight $m_\ell\rho^s/\gZ$ in
Equation~\ref{eq:retrace}. Both target types use the same heads and
supervised horizons. The future hard-SID control thus changes the
target-dependent gradient while holding the supervision locations fixed.

These derivatives specify how residual preferences affect the auxiliary
update. The training coefficient scales this contribution when the
trajectory and retrieval objectives are optimized jointly.

\subsection{Normalized Horizon Weights}
\label{app:horizon-weight-allocation}

With $L=4$, $H=4$, and all target positions valid, offset $s$ occurs
$4-s$ times in Equation~\ref{eq:retrace}. Its total normalized contribution
across all valid state--horizon pairs is
\begin{equation}
w_s=\frac{(4-s)\rho^s}{\sum_{u=0}^{3}(4-u)\rho^u}.
\label{eq:normalized-horizon-weight}
\end{equation}
For $\rho=0.7$, the unnormalized weights are
$(4,2.1,0.98,0.343)$ and $\gZ=7.423$. Hence
$(w_0,w_1,w_2,w_3)=(0.5389,0.2829,0.1320,0.0462)$.
With $H=1$, all weight is assigned to the current level. Increasing the
horizon therefore changes both the supervised targets and their relative
weights. RBD + future hard SID preserves the $H=4$ weight allocation and
changes only the future target distributions. These fractions are weights
in the objective. Realized loss and gradient contributions also depend on
the KL terms, model Jacobians, and loss scaling.

%% file: sections/appendix_algorithm.tex
\Needspace{27\baselineskip}
\subsection{Training Algorithm}
\label{app:algorithm}

\begin{algorithm}[H]
\caption{ResTD training with a frozen SID indexer}
\label{alg:retrace-training}
\begin{algorithmic}[1]
\Require Pairs $(q,d)$, stored SIDs $\vy(d)$, frozen codebooks
$\{\mathcal C_t\}_{t=1}^{L}$, horizon $H$, retriever $f_\theta$
\ForAll{training items $d$}
  \State Initialize $\vr_0\gets\vx_d$
  \For{$t=1,\ldots,L$}
    \State Precompute collision-corrected $\widetilde{\gQ}_t$ using Equation~\ref{eq:teacher} and Equations~\ref{eq:collision}--\ref{eq:collision_eps}
    \State $\vr_t\gets\vr_{t-1}-\vc_{t,y_t(d)}$
  \EndFor
\EndFor
\ForAll{optimization steps $u$ and minibatches of $(q,d)$}
  \State Compute original loss $\mathcal L_{\mathrm{GR}}$ and teacher-forced SID states $\vh_{1:L}$
  \For{$t=1,\ldots,L$ and $s=0,\ldots,\min(H,L-t+1)-1$}
    \State Evaluate $g_s(\vh_t)$ against $\mathcal C_{t+s}$ to obtain $\gP_{t,s}$
    \State Accumulate the weighted KL term in Equation~\ref{eq:retrace}
  \EndFor
  \State Update $\theta$ and $\{g_s\}$ using Equation~\ref{eq:total}; keep the indexer frozen
\EndFor
\State Discard teacher distributions and auxiliary heads at inference
\end{algorithmic}
\end{algorithm}

%% file: sections/appendix_computational_cost.tex
\subsection{Computational Complexity and Training Cost}\label{app:computational-cost}
\label{sec:rq8}

\subsubsection{Computational complexity}
\label{sec:complexity}

Each horizon head transforms the decoder state in dimension $d_h$ and
projects it to residual dimension $D$. The heads introduce
$H(d_h^2+d_hD+3d_h+D)$ parameters, independent of corpus size and codebook
cardinality. For a batch of size $B$, evaluating all valid codebook
distributions requires
$\mathcal{O}\!\left(BD\sum_{t=1}^{L}\sum_{s=0}^{H_t-1}K_{t+s}\right)$
operations. Distance computations can be parallelized over positions,
horizons, and codewords. Precomputing residual trajectories for $N$ items
requires $\mathcal{O}(N(L+1)D)$ storage. These computations and
trajectories are needed only during training, so they add no inference
cost. Appendix~\ref{app:training-overhead} reports measured training time,
GPU memory, and storage requirements.

\subsubsection{Measured training overhead}
\label{app:training-overhead}

Training cost is measured on ESCI-US with T5-base, BF16, AdamW, and one
A100-SXM4-80GB GPU. All three objectives use identical initial parameters,
minibatch order, a frozen category adapter, and four auxiliary heads.
Each update accumulates gradients over 16 microbatches of eight examples,
for an effective batch size of 128. Timing covers 12 optimizer updates
after three initial updates, with GPU synchronization at each update
boundary. Each measurement spans batch preparation and transfer,
forward and backward computation, gradient accumulation, and parameter
updates. The learning rate and auxiliary weight follow the schedules of
the 1,200-update experiments. All 12 measurements are taken during their
warmup phase.

In Table~\ref{tab:training-overhead-a100}, the update time for $H=4$ is
$1.119\times$ Base-FT and $1.057\times$ RBD ($H=1$), with 31.8\,MB
of additional peak allocated memory relative to Base-FT.
Across the 27 full-length experiments, comprising three locales, three
seeds, and three objectives with 1,200 updates each, the mean training
time for $H=4$ relative to Base-FT is $1.116\times$, $1.083\times$,
and $1.094\times$ on US, ES, and JP, respectively. The corresponding
ratios versus $H=1$ range from $1.060\times$ to $1.063\times$.
Full-run timings cover the optimization phase, from batch processing
through parameter updates.

Residual trajectories use FP16 precision with $D=32$ and $L=4$.
The US trajectories form an array of size $288{,}372\times5\times32$,
requiring 0.0923\,GB. Including SIDs, codebooks, and associated metadata
raises the total to 0.0947\,GB. The ES and JP trajectories require
0.0326 and 0.0381\,GB, respectively. The same precomputed trajectories
provide supervision for both $H=1$ and $H=4$.
\input{tables/table_training_overhead}

%% file: tables/table_training_overhead.tex
\begin{table}[!htbp]
\centering\appendixtablestyle
\caption{Training cost on ESCI-US with one A100 80GB GPU.
Time is mean \textpm{} sample standard deviation over 12 optimizer updates.
Peak GPU memory is the maximum allocation by PyTorch over all 15 updates,
including the three excluded from timing. GB denotes 10\textsuperscript{9} bytes.}
\label{tab:training-overhead-a100}
\begin{tabular*}{\linewidth}{@{\extracolsep{\fill}}lrrrrrl@{}}
\toprule
Objective & \shortstack{Time\\(s/update)} & \shortstack{Time\\vs. Base}
& \shortstack{Throughput\\(samples/s)} & \shortstack{GPU peak\\(GB)}
& \shortstack{Residual\\storage (GB)} & \shortstack{Inference\\overhead}\\
\midrule
Base-FT & 6.314\,\textpm\,0.024 & 1.000\texttimes{} & 20.27 & 4.837 & --\textsuperscript{*} & None\\
RBD (\textit{H}=1) & 6.686\,\textpm\,0.019 & 1.059\texttimes{} & 19.14 & 4.845 & 0.0923 & None\\
ResTD (\textit{H}=4) & 7.069\,\textpm\,0.018 & 1.119\texttimes{} & 18.11 & 4.869 & 0.0923 & None\\
\bottomrule
\end{tabular*}
\par\smallskip
\begin{minipage}{\linewidth}\footnotesize
\textsuperscript{*}Base-FT uses no residual supervision. For a controlled
comparison, it retains the same residual-data loading procedure and four
inactive auxiliary heads. Inference overhead refers to auxiliary computation.
\end{minipage}
\end{table}

%% file: sections/appendix_nq320k.tex
\subsection{Generalization to NQ320K}
\label{app:nq320k}

\paragraph{Dataset and evaluation setting.}
NQ320K extends our evaluation from product search to Wikipedia document
retrieval. Following \citet{sun2023learning}, we use 109,739 documents, 307,373 training
query--document pairs, and 7,830 test queries, each with one relevant
document. The test set comprises 6,075 seen and 1,755 unseen queries,
distinguished by whether their relevant documents appear in the training
pairs. Models train on the original pairs and retrieve from the full
corpus, with the unseen subset testing generalization to documents
outside the supervised training set.

\paragraph{Comparison protocol.}
We compare \method with TIGER, MERGE, and CaLIR. The baselines use
T5-base, trained for 100 epochs with AdamW, a learning rate of
$5\times10^{-4}$, and a per-device batch size of 512, with evaluation
at the final epoch. \method introduces residual trajectory supervision
into CaLIR training while preserving its SID index and decoding procedure.
All methods use trie-constrained beam search with beam size 100.
Recall@10, Recall@100, and MRR@100 measure retrieval coverage and ranking
quality across the full, seen, and unseen test sets.

\input{tables/table_nq320k}

\paragraph{Results and generalization.}
Table~\ref{tab:nq320k-results} shows that \method leads all three baselines
on every metric across the full, seen, and unseen test sets. Gains in
recall and reciprocal rank show that \method retrieves more relevant
documents and places them earlier in the ranking. On unseen documents,
Recall@10 and MRR@100 more than double the strongest baseline scores,
alongside higher \mbox{Recall@100}, demonstrating generalization beyond
supervised query--document associations. The gains on seen documents show
that improved generalization accompanies stronger retrieval of supervised
documents. NQ320K extends these benefits from product search with multiple
relevant items to natural-language questions with a single relevant
document per query. These results establish residual trajectory
supervision as an effective training signal across retrieval domains and
relevance settings.

%% file: tables/table_nq320k.tex
\begin{table}[!htbp]
\caption{Retrieval performance on NQ320K (\%). R@$K$ denotes Recall@$K$
and MRR denotes MRR@100. Bold and underlined values indicate the best
and second-best results, respectively. Improv. ($\Delta$) reports gains
in percentage points over the strongest baseline in each column.}
\label{tab:nq320k-results}
\centering\appendixtablestyle
\setlength{\tabcolsep}{2.5pt}
\begin{tabular*}{\linewidth}{@{\extracolsep{\fill}}l*{9}{r}@{}}
\toprule
& \multicolumn{3}{c}{\textbf{Full test}}
& \multicolumn{3}{c}{\textbf{Seen test}}
& \multicolumn{3}{c}{\textbf{Unseen test}}\\
\cmidrule(lr){2-4}\cmidrule(lr){5-7}\cmidrule(lr){8-10}
\textbf{Method} & R@10 & R@100 & MRR
& R@10 & R@100 & MRR
& R@10 & R@100 & MRR\\
\midrule
TIGER & 53.72 & 60.20 & 47.22
& 68.86 & 74.65 & 60.71
& 1.31 & 10.20 & 0.54\\
MERGE & 54.23 & 61.28 & 47.07
& 69.30 & 75.28 & 60.44
& \underline{2.05} & 12.82 & \underline{0.77}\\
CaLIR & \underline{56.54} & \underline{63.75} & \underline{51.51}
& \underline{72.72} & \underline{76.21} & \underline{66.24}
& 0.51 & \underline{20.63} & 0.51\\
\textbf{\method} & \textbf{59.41} & \textbf{65.75} & \textbf{53.86}
& \textbf{75.19} & \textbf{77.23} & \textbf{68.83}
& \textbf{4.79} & \textbf{25.98} & \textbf{2.04}\\
\midrule
\emph{Improv.} ($\Delta$) & +2.87 & +2.00 & +2.35
& +2.47 & +1.02 & +2.59
& +2.74 & +5.35 & +1.27\\
\bottomrule
\end{tabular*}
\end{table}

%% file: sections/appendix_teacher_geometry.tex
\subsection{Teacher-Target Controls}
\label{app:teacher-geometry-controls}

\subsubsection{Within-Code Variation in Teacher Distributions}
\label{app:within-code-variation}
\label{sec:teacher-information}

\begin{figure}[!htbp]
  \centering
  \includegraphics[width=0.60\linewidth]{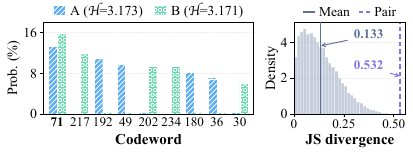}
  \caption{Within-code teacher variation on ESCI-US. Left: an illustrative
  pair selected for high divergence. Right: JS divergence over 20,000 pairs
  matched by hard code and entropy decile. Solid and dashed lines mark the
  mean and the example, respectively.}
  \label{fig:residual-variation}
\end{figure}

We examine 20,000 ESCI-US item pairs with the same first-level hard code
and teacher-entropy decile. Their mean Jensen--Shannon (JS) divergence is
0.1330 nats, with a 95th percentile of 0.3088
(Figure~\ref{fig:residual-variation}). The illustrated pair is selected
for high divergence: both items are assigned to $c_{71}$ and have similar
entropies, 3.173 and 3.171 nats, yet their strongest alternatives are
$c_{192}$ and $c_{217}$. The JS divergence between their distributions is
0.5322 nats. This variation captures information beyond the hard code.
Section~\ref{sec:teacher-geometry-controls} evaluates its contribution
to retrieval performance.

\subsubsection{Additional Retrieval Results}
Table~\ref{tab:teacher-geometry-full} extends the comparison in
Section~\ref{sec:teacher-geometry-controls} to all five retrieval metrics,
including Recall@5. Entries are means and sample standard deviations
across three training seeds for each objective and locale.

Table~\ref{tab:teacher-properties} summarizes the control properties using
the definitions in Table~\ref{tab:teacher-geometry-controls}.
\emph{Hard} denotes preservation of the item's assigned hard code.
\emph{Shape} describes how the target preserves the original teacher's
permutation-invariant probability profile: \emph{agg.} denotes aggregate
matching, \emph{coarse} entropy-bin matching with a singleton fallback,
\emph{near} approximate shape
matching, and \emph{exact} preservation of the probability profile;
\emph{unmatched} denotes unrestricted teacher shuffling.
\emph{Real} denotes an unmodified distribution induced by an actual item.
Averages and probability permutations are transformed targets.
\emph{Pair} denotes preservation of the original item--teacher correspondence.
Checkmarks and crosses indicate whether a binary property is preserved.
Dashes denote inapplicable properties. Base-FT has no auxiliary teacher.
For RBD + future hard SID, the properties refer to the hard future
targets, while current-level supervision remains unchanged.
For Codebook-only ($H=4$), targets are constructed from the stored code and
frozen codebook without item residuals or corpus residual statistics.
The dash in Shape indicates that no residual-teacher probability profile
is used or matched.

\input{tables/table_teacher_properties}
\input{tables/table_teacher_geometry_full}

\subsubsection{Control Construction}
\label{app:teacher-control-protocol}

Let $\mathcal D_{\mathrm{train}}$ be the set of distinct training items.
For an item $d$ and target level $\ell$, write $y_\ell(d)$ for its stored
code and $\widetilde{\gQ}_\ell(d)$ for its collision-corrected teacher.
Each control defines a teacher for a valid item--level pair $(d,\ell)$
before training. This teacher is held fixed and used for every valid
state--horizon pair $(t,s)$ satisfying $t+s=\ell$. The transformation
therefore applies to both current ($s=0$) and future ($s>0$) targets.
Table~\ref{tab:teacher-control-protocol} summarizes these constructions.
RBD + future hard SID is the exception: it preserves current-level RBD
and replaces only $s>0$ targets with one-hot identifiers.

\paragraph{Non-target permutation.}
For each $(d,\ell)$, a fixed random permutation reassigns the probabilities
of the $K_\ell-1$ non-target codewords while leaving
$\widetilde{\gQ}_\ell(y_\ell(d)\mid d)$ unchanged. The complete probability
multiset is preserved, including entropy, assigned-code confidence, and
the largest alternative probability. Only the association between
non-target codewords and their probabilities changes. This permutation
is held fixed wherever the item--level target is used.

\paragraph{Codebook-only targets.}
For each target level $\ell$, we construct a distribution from the distances
between the stored codeword and all codewords in the frozen codebook.
Items sharing a stored code at the same level receive identical targets.
This uses no item residuals or corpus residual statistics.

\paragraph{Code-conditioned mean.}
For each target level $\ell$ and stored code $c$, define
\begin{equation}
\overline{\gQ}_{\ell,c}
=\frac{1}{|\mathcal D_{\ell,c}|}
\sum_{d\in\mathcal D_{\ell,c}}\widetilde{\gQ}_\ell(d),
\qquad
\mathcal D_{\ell,c}
=\{d\in\mathcal D_{\mathrm{train}}:y_\ell(d)=c\}.
\label{eq:code-conditioned-teacher}
\end{equation}
The mean is computed before training for each level and code, with equal
weight for every distinct item. Assigning this distribution to all items
in $\mathcal D_{\ell,c}$ preserves code-conditioned population preferences
while removing variation among items with the same code.

\paragraph{Same-code + entropy-bin shuffle.}
At each target level, we stably sort the corrected training teachers by
$(\mathcal H(\widetilde{\gQ}_\ell(d)),\mathrm{item\ ID})$ and partition them
into ten equal-frequency entropy bins. Within each group sharing the
same level, stored hard code, and entropy bin, items are ordered by ID
and their teachers are cyclically permuted. Groups of size at least two
therefore receive a teacher from another item. For a singleton group, the donor is
the other training item with the closest teacher entropy at the same
level and stored code; ties are resolved by item ID. This fallback may
cross an entropy-bin boundary while preserving the hard code.
Assignments remain fixed throughout training. The 100\% replacement rate
includes these singleton assignments and therefore measures replacement
by another item's teacher, not exact entropy-bin agreement.

\paragraph{Shape-matched teacher swap.}
For a corrected teacher $\widetilde{\gQ}$ with stored code $c$, define
\begin{equation}
v_c(\widetilde{\gQ})=\left[\widetilde{\gQ}(c),\,
\operatorname{sort}_{\downarrow}
\{\widetilde{\gQ}(k):k\ne c\}\right].
\label{eq:teacher-shape-vector}
\end{equation}
Candidate donors are other training items at the same target level and
stored hard code. We use the $L_1$ distance between shape vectors for
greedy one-to-one nearest matching without replacement, forming disjoint
item pairs. Distance ties are resolved deterministically by item ID.
The two items in each pair exchange their complete, unmodified corrected
teachers, so no donor is reused. In a group of odd size, the final
unpaired item retains its own teacher and is counted as unswapped.
The pairing is constructed once before training and shared across
optimization steps and training seeds 42, 2027, and 2028. The replacement
rate of 99.76--99.91\% excludes items that retain their own teacher.

The replacement rates in Table~\ref{tab:teacher-control-protocol} count
transformation applications for synthetic, aggregate, and hard targets,
including those that leave the target distribution unchanged. For real-item
controls, they count assignments to another item's teacher. Each rate uses
all valid item--level targets in the indicated scope. This ensures consistent
accounting across the different control constructions.

\input{tables/table_teacher_control_protocol}

\subsubsection{Matching Quality}
\label{app:shape-matching-quality}

For a donor--recipient pair $(d',d)$ at level $\ell$, define
$\gQ'=\widetilde{\gQ}_\ell(d')$, $\gQ=\widetilde{\gQ}_\ell(d)$, and
$c=y_\ell(d')=y_\ell(d)$. We quantify entropy, assigned-code probability,
and shape differences as
\begin{align}
|\Delta\mathcal H|&=|\mathcal H(\gQ')-\mathcal H(\gQ)|,
&\mathcal H(\gQ)&=-\sum_k \gQ(k)\log \gQ(k),\nonumber\\
|\Delta q_y|&=|\gQ'(c)-\gQ(c)|,
& D_{\mathrm{shape}}&=\|v_c(\gQ')-v_c(\gQ)\|_1.
\label{eq:teacher-matching-quality}
\end{align}
Entropy is measured in nats. Matching statistics are computed over
donor--recipient pairs with distinct item identities. Coverage is the
fraction of valid training item--level targets assigned another item's
teacher, even when the donor and recipient have identical probability vectors.

\input{tables/table_shape_matching_quality}

Median entropy mismatch is 0.008--0.012 nats, and median assigned-code
probability mismatch is 0.0018--0.0027, corresponding to 0.18--0.27
percentage points. Median Shape $L_1$ is 0.018--0.026, with P90 values
of 0.046--0.064. Different-item coverage is 99.76--99.91\%.
The label \emph{near} in Table~\ref{tab:teacher-geometry-controls} refers to
this approximate shape matching. Similar profiles can have different
codeword probabilities.

%% file: tables/table_teacher_properties.tex
\begin{table}[!htbp]
\centering\appendixtablestyle
\caption{Properties of the teacher-target controls. Hard, Shape, Real, and Pair
follow the definitions in Table~\ref{tab:teacher-geometry-controls}.
Properties are shared across all three locales.}
\label{tab:teacher-properties}
\begin{tabularx}{\linewidth}{@{}Xcccc@{}}
\toprule
Teacher objective & Hard & Shape & Real & Pair\\
\midrule
Base-FT & $\checkmark$ & -- & -- & --\\
Shuffled ResTD & \texttimes{} & unmatched & $\checkmark$ & \texttimes{}\\
RBD + future hard SID & $\checkmark$ & -- & -- & --\\
Codebook-only (\textit{H}=4) & $\checkmark$ & -- & \texttimes{} & \texttimes{}\\
Code-conditioned mean & $\checkmark$ & agg. & \texttimes{} & \texttimes{}\\
Same-code + entropy-bin shuffle & $\checkmark$ & coarse & $\checkmark$ & \texttimes{}\\
Non-target permutation & $\checkmark$ & exact & \texttimes{} & \texttimes{}\\
Shape-matched teacher swap & $\checkmark$ & near & $\checkmark$ & \texttimes{}\\
\rowcolor{mainresultrow}\textbf{ResTD (\textit{H}=4)} & $\checkmark$ & exact & $\checkmark$ & $\checkmark$\\
\bottomrule
\end{tabularx}
\end{table}

%% file: tables/table_teacher_geometry_full.tex
\begin{table}[!htbp]
\centering\appendixtablestyle
\caption{Complete teacher-target comparison on ESCI (\%), including Recall@5. Entries are mean \textpm{} standard deviation across three training seeds. Teacher properties and full control names are given in Table~\ref{tab:teacher-properties}. Bold marks the highest means.}
\label{tab:teacher-geometry-full}
\begin{tabularx}{\linewidth}{@{}L{.32\linewidth}*{5}{>{\raggedleft\arraybackslash}X}@{}}
\toprule
Teacher objective & R@5 & R@10 & R@100 & N@10 & N@100\\
\midrule
\multicolumn{6}{l}{\textbf{ESCI-US}}\\
\addlinespace[2pt]
Base-FT & \appendixstat{7.62}{0.03} & \appendixstat{12.82}{0.16} & \appendixstat{37.60}{0.13} & \appendixstat{11.25}{0.11} & \appendixstat{18.92}{0.15}\\
Shuffled ResTD & \appendixstat{7.65}{0.14} & \appendixstat{12.86}{0.05} & \appendixstat{37.68}{0.11} & \appendixstat{11.22}{0.12} & \appendixstat{18.95}{0.03}\\
RBD + future hard SID & \appendixstat{8.51}{0.16} & \appendixstat{13.56}{0.09} & \appendixstat{38.68}{0.07} & \appendixstat{12.18}{0.08} & \appendixstat{19.93}{0.13}\\
Codebook-only (\textit{H}=4) & \appendixstat{8.38}{0.10} & \appendixstat{13.48}{0.11} & \appendixstat{38.55}{0.06} & \appendixstat{12.05}{0.14} & \appendixstat{19.81}{0.03}\\
Code-conditioned mean & \appendixstat{8.91}{0.12} & \appendixstat{13.90}{0.08} & \appendixstat{39.03}{0.15} & \appendixstat{12.50}{0.10} & \appendixstat{20.27}{0.16}\\
Entropy-bin shuffle & \appendixstat{8.88}{0.04} & \appendixstat{13.82}{0.07} & \appendixstat{38.96}{0.10} & \appendixstat{12.44}{0.06} & \appendixstat{20.22}{0.11}\\
Non-target permutation & \appendixstat{8.73}{0.12} & \appendixstat{13.76}{0.10} & \appendixstat{38.86}{0.16} & \appendixstat{12.36}{0.15} & \appendixstat{20.12}{0.09}\\
Shape-matched swap & \appendixstat{9.13}{0.17} & \appendixstat{14.16}{0.10} & \appendixstat{39.36}{0.07} & \appendixstat{12.79}{0.03} & \appendixstat{20.58}{0.11}\\
\rowcolor{mainresultrow}\textbf{ResTD (\textit{H}=4)} & \appendixbeststat{9.28}{0.09} & \appendixbeststat{14.34}{0.04} & \appendixbeststat{39.58}{0.11} & \appendixbeststat{12.99}{0.07} & \appendixbeststat{20.79}{0.05}\\
\midrule
\multicolumn{6}{l}{\textbf{ESCI-ES}}\\
\addlinespace[2pt]
Base-FT & \appendixstat{6.10}{0.10} & \appendixstat{10.55}{0.15} & \appendixstat{34.22}{0.11} & \appendixstat{12.53}{0.12} & \appendixstat{19.95}{0.09}\\
Shuffled ResTD & \appendixstat{6.08}{0.11} & \appendixstat{10.51}{0.06} & \appendixstat{34.10}{0.04} & \appendixstat{12.60}{0.14} & \appendixstat{19.92}{0.03}\\
RBD + future hard SID & \appendixstat{6.47}{0.09} & \appendixstat{11.16}{0.04} & \appendixstat{34.81}{0.05} & \appendixstat{13.40}{0.10} & \appendixstat{20.71}{0.15}\\
Codebook-only (\textit{H}=4) & \appendixstat{6.35}{0.09} & \appendixstat{10.98}{0.11} & \appendixstat{34.65}{0.13} & \appendixstat{13.22}{0.12} & \appendixstat{20.51}{0.08}\\
Code-conditioned mean & \appendixstat{6.62}{0.15} & \appendixstat{11.39}{0.06} & \appendixstat{35.00}{0.08} & \appendixstat{13.66}{0.03} & \appendixstat{20.97}{0.13}\\
Entropy-bin shuffle & \appendixstat{6.50}{0.14} & \appendixstat{11.29}{0.11} & \appendixstat{34.89}{0.07} & \appendixstat{13.60}{0.09} & \appendixstat{20.91}{0.12}\\
Non-target permutation & \appendixstat{6.55}{0.03} & \appendixstat{11.29}{0.04} & \appendixstat{34.93}{0.15} & \appendixstat{13.54}{0.12} & \appendixstat{20.84}{0.06}\\
Shape-matched swap & \appendixstat{6.76}{0.08} & \appendixstat{11.61}{0.11} & \appendixstat{35.23}{0.03} & \appendixstat{14.00}{0.05} & \appendixstat{21.31}{0.07}\\
\rowcolor{mainresultrow}\textbf{ResTD (\textit{H}=4)} & \appendixbeststat{6.86}{0.03} & \appendixbeststat{11.76}{0.04} & \appendixbeststat{35.38}{0.13} & \appendixbeststat{14.22}{0.07} & \appendixbeststat{21.54}{0.12}\\
\midrule
\multicolumn{6}{l}{\textbf{ESCI-JP}}\\
\addlinespace[2pt]
Base-FT & \appendixstat{6.08}{0.13} & \appendixstat{9.85}{0.03} & \appendixstat{30.98}{0.07} & \appendixstat{11.90}{0.17} & \appendixstat{18.55}{0.08}\\
Shuffled ResTD & \appendixstat{6.10}{0.04} & \appendixstat{9.89}{0.06} & \appendixstat{31.02}{0.12} & \appendixstat{11.88}{0.13} & \appendixstat{18.59}{0.07}\\
RBD + future hard SID & \appendixstat{6.78}{0.12} & \appendixstat{10.81}{0.03} & \appendixstat{31.96}{0.11} & \appendixstat{13.12}{0.07} & \appendixstat{19.72}{0.16}\\
Codebook-only (\textit{H}=4) & \appendixstat{6.62}{0.10} & \appendixstat{10.63}{0.14} & \appendixstat{31.76}{0.09} & \appendixstat{12.95}{0.15} & \appendixstat{19.51}{0.16}\\
Code-conditioned mean & \appendixstat{7.11}{0.08} & \appendixstat{11.09}{0.13} & \appendixstat{32.24}{0.10} & \appendixstat{13.47}{0.05} & \appendixstat{20.05}{0.15}\\
Entropy-bin shuffle & \appendixstat{7.09}{0.08} & \appendixstat{11.03}{0.04} & \appendixstat{32.17}{0.13} & \appendixstat{13.44}{0.10} & \appendixstat{20.03}{0.05}\\
Non-target permutation & \appendixstat{6.96}{0.17} & \appendixstat{10.96}{0.04} & \appendixstat{32.11}{0.05} & \appendixstat{13.31}{0.10} & \appendixstat{19.89}{0.12}\\
Shape-matched swap & \appendixstat{7.41}{0.12} & \appendixstat{11.41}{0.04} & \appendixstat{32.61}{0.07} & \appendixstat{13.90}{0.09} & \appendixstat{20.49}{0.17}\\
\rowcolor{mainresultrow}\textbf{ResTD (\textit{H}=4)} & \appendixbeststat{7.61}{0.07} & \appendixbeststat{11.63}{0.17} & \appendixbeststat{32.86}{0.09} & \appendixbeststat{14.18}{0.08} & \appendixbeststat{20.79}{0.05}\\
\bottomrule
\end{tabularx}
\end{table}

%% file: tables/table_teacher_control_protocol.tex
\begin{table}[!htbp]
\centering\appendixtablestyle
\caption{Construction of teacher controls across current and future targets.
Assignments remain fixed during training; replacement rates are measured
over valid item--level targets within each scope.}
\label{tab:teacher-control-protocol}
\begin{tabularx}{\linewidth}{@{}p{.235\linewidth}lp{.265\linewidth}cX@{}}
\toprule
Control & Scope & Matching unit & Fixed & Target replacement\\
\midrule
Non-target permutation & Current + future & Item \texttimes{} level & $\checkmark$ & 100\% synthetic\\
Code-conditioned mean & Current + future & Level \texttimes{} hard code & $\checkmark$ & 100\% aggregate\\
Entropy-bin shuffle & Current + future & Level \texttimes{} code \texttimes{} entropy decile\textsuperscript{*} & $\checkmark$ & 100\% real-item\\
Shape-matched swap & Current + future & Level \texttimes{} code & $\checkmark$ & 99.76--99.91\% real-item\\
Future hard SID & Future only & -- & $\checkmark$ & 100\% hard target\\
\bottomrule
\end{tabularx}
\par\smallskip
\begin{minipage}{\linewidth}\footnotesize
\textsuperscript{*}Singleton groups retain the level and hard code but may use a donor from another entropy bin.
\end{minipage}
\end{table}

%% file: tables/table_shape_matching_quality.tex
\begin{table}[!htbp]
\centering\appendixtablestyle
\caption{Matching quality for shape-matched teacher swap. Mismatch summaries
are medians except P90 (90th percentile), computed over pairs with different
item identities. Coverage uses all valid training item--level targets.
Entropy uses nats; probability differences and Shape \textit{L}\textsubscript{1} are unitless.}
\label{tab:shape-matching-quality}
\begin{tabular*}{\linewidth}{@{\extracolsep{\fill}}lrrrrr@{}}
\toprule
 & \shortstack{Median\\$|\Delta\mathcal H|$}
 & \shortstack{Median\\$|\Delta q_y|$}
 & \multicolumn{2}{c}{Shape \textit{L}\textsubscript{1}}
 & \shortstack{Swap\\coverage (\%)}\\
\cmidrule(lr){4-5}
Locale & (nats) & & Median & P90 & \\
\midrule
US & 0.008 & 0.0018 & 0.018 & 0.046 & 99.91\\
ES & 0.012 & 0.0027 & 0.026 & 0.064 & 99.76\\
JP & 0.010 & 0.0023 & 0.023 & 0.057 & 99.81\\
\bottomrule
\end{tabular*}
\end{table}

%% file: sections/appendix_output_supervision.tex
\subsection{Output-Space Supervision}
\label{app:output-supervision}
\label{sec:rq8-attribution}

\subsubsection{Full-Distribution Output Distillation}
\label{app:full-output-kd}

Full-output KD applies the collision-corrected current-level teacher
$\widetilde{\gQ}_t$ directly to the retriever's SID output distribution.
Let $z_{t,j}$ denote its output logit for codeword $j$ at level $t$. We define
\begin{equation}
\gP_t^{\mathrm{out}}(j\mid q,y_{<t})
=\frac{\exp z_{t,j}}{\sum_{k=1}^{K_t}\exp z_{t,k}},
\qquad
\mathcal L_{\mathrm{FullOut}}
=\frac{\sum_{t=1}^{L}m_t\,\mathrm{KL}
(\widetilde{\gQ}_t\|\gP_t^{\mathrm{out}})}{\sum_{t=1}^{L}m_t}.
\label{eq:full-output-kd}
\end{equation}
Here $m_t$ selects valid SID positions. The full $K_t=256$-way teacher
retains the probabilities of the assigned code and all alternatives.
Supervision targets the current SID position.

Full-output KD and RBD ($H=1$) use the same teacher and train for 1,200
updates with learning rate $10^{-5}$ and 120-step warmup. Initial parameters
and minibatch order are matched within seeds 42, 2027, and 2028. Both
retain the original retrieval objective and use the auxiliary schedule in
Equation~\ref{eq:total}, with $\lambda_{\max}=0.1$ and a cap of 5\% of the
SID loss, treated as constant during backpropagation.
Full-output KD supervises SID output probabilities. RBD supervises a
projected decoder state evaluated against the frozen codebook. Gradient
norms depend on the supervision parameterization.

\input{tables/table_full_output_kd}

Full-output KD improves over Base-FT by 0.50, 0.18, and 0.41 percentage
points in Recall@100 on US, ES, and JP, respectively. Current-level RBD
adds 0.10, 0.06, and 0.09 points over Full-output KD, with NDCG@10 gains
of 0.09, 0.07, and 0.11 points. Both objectives improve retrieval,
with higher mean scores for current-level codebook distillation.
\subsubsection{Non-Target Output Alignment}
\label{app:nontarget-output-alignment}
This control aligns the retriever's output probabilities over
alternative codes. At each valid position, we remove the assigned code from the
teacher and student distributions and renormalize the remaining mass.
Writing these conditional distributions as $\overline{\gQ}_t$ and
$\overline{\gP}_t$, the auxiliary loss is
\begin{equation}
\mathcal L_{\mathrm{out}}
=\frac{1}{|\mathcal T|}\sum_{t\in\mathcal T}
\mathrm{KL}(\overline{\gQ}_t\|\overline{\gP}_t),
\label{eq:output-alignment}
\end{equation}
where $\mathcal T$ contains valid positions with positive alternative teacher mass.
The original gold-SID objective is unchanged. The auxiliary term has
coefficient 0.02 and is capped at 1\% of the SID loss. The cap is treated
as constant during backpropagation. Table~\ref{tab:core-training}
reports the retrieval results.

%% file: tables/table_full_output_kd.tex
\begin{table}[!htbp]
\centering\appendixtablestyle
\caption{Full-output KD on ESCI (\%). Values are mean \textpm{} sample standard
 deviation across seeds 42, 2027, and 2028. All five retrieval metrics are reported.}
\label{tab:full-output-kd}
\begin{tabular*}{\linewidth}{@{\extracolsep{\fill}}lrrrrr@{}}
\toprule
Locale & R@5 & R@10 & R@100 & N@10 & N@100\\
\midrule
US & \appendixstat{7.95}{0.08} & \appendixstat{13.15}{0.07} & \appendixstat{38.10}{0.13} & \appendixstat{11.69}{0.10} & \appendixstat{19.35}{0.09}\\
ES & \appendixstat{6.27}{0.07} & \appendixstat{10.80}{0.05} & \appendixstat{34.40}{0.11} & \appendixstat{12.91}{0.06} & \appendixstat{20.20}{0.14}\\
JP & \appendixstat{6.35}{0.08} & \appendixstat{10.22}{0.09} & \appendixstat{31.39}{0.08} & \appendixstat{12.39}{0.11} & \appendixstat{18.99}{0.12}\\
\bottomrule
\end{tabular*}
\end{table}

%% file: sections/appendix_future_probe.tex
\subsection{Probing Current and Future Codebook Preferences}
\label{app:future-probe}

\subsubsection{Current-Target Diagnostic}
\label{app:current-target-probe}

\begin{figure}[!htbp]
  \centering
  \includegraphics[width=0.65\linewidth]{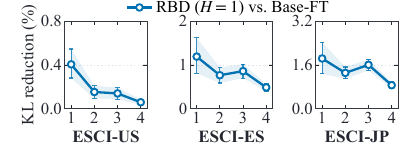}
  \caption{Relative held-out probe KL reduction for current-level RBD
  ($H=1$) versus Base-FT across four SID levels. Error bars show paired
  95\% query-bootstrap intervals conditional on one pair of trained
  Base-FT and RBD models for each locale.}
  \label{fig:process-transfer}
\end{figure}

For the current-level diagnostic in Figure~\ref{fig:process-transfer}, we
freeze one pair of trained Base-FT and RBD models per locale and fit a new
projection for each model. The projection is shared across all four SID
levels and consists of a linear map from $d_h$ to $d_h$,
GELU, LayerNorm, and a linear map to residual dimension $D$. It contains
$d_h^2+d_hD+3d_h+D$ trainable parameters. The probes are fitted for ten
epochs with AdamW, learning rate $10^{-3}$, and weight decay 0.01.
Probe initialization and minibatch order use seed 4242 for both retrievers
within each locale. We evaluate the epoch-10 probes.

The fitting sets contain 991, 267, and 319 queries for US, ES, and JP,
respectively. Each evaluation set contains 256 disjoint queries. Targets
are collision-corrected teachers with floor 0.1 and margin 0.001.
We compute paired query-bootstrap intervals for each fixed pair of
trained retrievers.

Figure~\ref{fig:process-transfer} reports relative KL reductions of
0.06--1.84\% for current-level RBD versus Base-FT across the three locales
and four SID levels. This diagnostic measures recovery of current-level
teacher preferences from the frozen decoder representations.

\subsubsection{Future-Target Comparison}
\label{app:future-target-protocol}

We compare frozen RBD ($H=1$) and ResTD ($H=4$) models.
For each retriever, we fit a separate probe at each offset
$s=j-t\in\{0,1,2,3\}$. All valid state--target pairs $(t,j)$ with the
same offset share that probe. Each probe has the architecture
\begin{equation}
\mathrm{Linear}(d_h,d_h)\ \to\ \mathrm{GELU}\ \to\
\mathrm{LayerNorm}\ \to\ \mathrm{Linear}(d_h,D).
\label{eq:future-probe-architecture}
\end{equation}
With $d_h=768$ and $D=32$, each probe contains
$d_h^2+d_hD+3d_h+D=616{,}736$ trainable parameters. The probe capacity
and fitting procedure are identical across retriever conditions.

For each locale, 1,024 queries are used to fit the probes and 512 disjoint
queries are reserved for evaluation. These splits are fixed across
conditions and retriever seeds. Probes are trained for ten epochs with
AdamW, learning rate $10^{-3}$, weight decay 0.01, and batch size 64.
We evaluate the epoch-10 probes.
Probe initialization and minibatch order use seed 4242 throughout.
The three replicates correspond to retrievers trained with seeds 42,
2027, and 2028, with probe initialization and query splits held fixed.

The target is the collision-corrected teacher $\widetilde{\gQ}_j$ used in
retriever training, with floor 0.1 and margin 0.001. We report held-out
forward KL from the complete corrected teacher, including the assigned-code
mass, to the probe's codebook distribution.

The two probing studies use separate query splits. Model comparisons
use the common split within each study.
For a four-level SID, offset $s$ has $4-s$ valid state--target pairs.
Although pairs with the same offset share probe parameters, held-out KL
is evaluated separately for each valid pair $(t,j)$.
Figure~\ref{fig:future-probe} reports the relative reduction
$100(1-\mathrm{KL}_{H=4}(t,j)/\mathrm{KL}_{H=1}(t,j))$ from current-level
RBD to ResTD. Diagonal cells ($j=t$) correspond to current targets;
cells with $j>t$ correspond to future targets.

%% file: sections/appendix_multipositive.tex
\subsection{Multi-Positive Teacher Disagreement}
\label{app:multipositive-details}

\subsubsection{Item-Conditioned Teacher Targets}

Several relevant items can induce different teachers for the same query.
Let $\mathcal D^+(q,p)$ contain the relevant items whose SIDs share prefix
$p$ before position $t$. They share the teacher-forced decoding context, but
may induce different codebook distributions $\gQ_t(d)$. Under uniform sampling
within this set, write $\overline{\gQ}_t=\mathbb E_d \gQ_t(d)$. For any common
student distribution $\gP_t$, the forward-KL decomposition is
\begin{equation}
\mathbb E_d\,\mathrm{KL}(\gQ_t(d)\|\gP_t)
=\mathbb E_d\,\mathrm{KL}(\gQ_t(d)\|\overline{\gQ}_t)
+\mathrm{KL}(\overline{\gQ}_t\|\gP_t).
\label{eq:teacher-ambiguity}
\end{equation}
For a shared decoding context, an unrestricted student minimizes the
expected forward-KL objective at the mean teacher distribution
$\overline{\gQ}_t$. The first term is the generalized Jensen--Shannon
divergence among the item-conditioned teachers and is constant with respect
to the student. With nonuniform item sampling, the identity holds for
the corresponding weighted mean.

To obtain Equation~\ref{eq:teacher-ambiguity}, expand the KL terms and
collect the coefficient of
$\log(\overline{\gQ}_t/\gP_t)$. Since $\mathbb E_d \gQ_t(d)=\overline{\gQ}_t$,
this coefficient is $\overline{\gQ}_t$. The remaining term equals
$H(\overline{\gQ}_t)-\mathbb E_d H(\gQ_t(d))$, the generalized JS divergence.
This decomposition concerns teachers at a common decoding context.
The tables below report mean pairwise JS, which is bounded by
$\log 2$. Generalized JS over several teachers need not satisfy that bound.

\input{tables/table_disagreement_strata_gains}

\subsubsection{Teacher Disagreement Statistics}

We compute disagreement from the first-level teachers used in training,
including the SID-consistency correction. For each query with at least
two distinct E/S/C-positive items, we average the pairwise JS divergence
between its positive-item teachers. Dataset means assign equal weight
to each query.

US, ES, and JP contain 4,649, 1,348, and 1,534 multi-positive queries
out of 6,014, 1,656, and 1,883 test queries, respectively. Their mean
JS divergences are 0.389, 0.557, and 0.543 nats. The remaining 1,365,
308, and 349 queries have a single positive item, respectively.
Within the multi-positive sets, 5 US, 12 ES, and 1 JP queries contain
distinct positive items with identical complete SIDs. Disagreement uses
distinct items, and retrieval metrics use distinct SIDs.
\subsubsection{Retrieval Gains by Teacher Disagreement}
Within each locale, we divide multi-positive queries into low, medium,
and high first-level disagreement groups of approximately equal size.
For each query, we compare ResTD ($H=4$) with Base-FT and current-level
RBD ($H=1$), averaging the paired metric differences across three runs.
Differences are reported in percentage points, and each query receives
equal weight within its group.
Table~\ref{tab:disagreement-strata-gains} reports query counts, mean
pairwise JS divergence in nats, mean positive-item counts, and the five
retrieval gains. The group sizes differ by at most one query.
Figure~\ref{fig:disagreement-gains} plots the group means for Recall@100
and NDCG@10. The stratified comparisons and adjusted regressions examine
query-level associations between teacher disagreement and retrieval gains.

\input{tables/table_adjusted_retrieval_associations}

\subsubsection{Adjusted Associations}
\label{app:retrieval-associations}
Let $J_q$ denote first-level teacher disagreement and $m_q$ the number
of distinct positive items. Initial query difficulty is measured using
our reproduced CaLIR model $\theta_{\mathrm{init}}$, before fine-tuning
with Base-FT, RBD, or ResTD. For the metric $M$ under analysis, define
\begin{equation}
s_q^{(M)}=M(q;\theta_{\mathrm{init}}),
\qquad
M\in\{\mathrm{Recall}@100,\mathrm{NDCG}@10\}.
\label{eq:initial-query-score}
\end{equation}
Initial and final scores use the same decoder and SID-level evaluation
protocol (Appendix~\ref{app:sid-evaluation}). The initial score is fixed
for each query and metric across all conditions and seeds. Both initial
and final scores are expressed on the $[0,100]$ scale. The paired gain over
baseline $b\in\{\text{Base-FT},H=1\}$ is
\begin{equation}
\overline\Delta_{q,b}^{(M)}
=\frac{1}{3}\sum_{r=1}^{3}
\left[M(q;\theta_{H4,r})-M(q;\theta_{b,r})\right],
\label{eq:paired-query-gain}
\end{equation}
where $r$ indexes the three paired runs. For each locale, metric, and
baseline, we fit
\begin{equation}
  \overline\Delta_{q,b}^{(M)}
  = \alpha + \beta\frac{J_q}{0.1}
    + \gamma\log(1+m_q) + \eta s_q^{(M)} + \varepsilon_q,
  \label{eq:adjusted-retrieval-association}
\end{equation}
where $\overline\Delta_{q,b}^{(M)}$ is the mean gain of ResTD over baseline
$b$, in percentage points across the three paired runs. Each multi-positive
query contributes one observation. The coefficient $\beta$ measures the
adjusted change in gain per
0.1-nat increase in JS. Table~\ref{tab:adjusted-retrieval-associations}
reports the raw rank correlations and coefficients with 95\% HC3
heteroskedasticity-robust confidence intervals.
The intervals quantify query-level uncertainty conditional on the fitted
retrievers.

For each 0.1-nat increase in JS, the adjusted NDCG@10 gain over $H=1$
is 0.17 points (95\% CI: 0.08--0.25) on US, 0.06 points
($-0.19$--0.31) on ES, and 0.62 points (0.39--0.85) on JP.

%% file: tables/table_disagreement_strata_gains.tex
\begin{table}[!htb]
\centering\appendixtablestyle
\renewcommand{\arraystretch}{0.98}
\caption{ResTD (\textit{H}=4) gains by first-level teacher disagreement among multi-positive queries. JS is mean pairwise Jensen--Shannon divergence in nats; Pos./query is the mean number of distinct positive items. Metric differences are in percentage points, averaged across three runs.}
\label{tab:disagreement-strata-gains}
\begin{tabular*}{\linewidth}{@{\extracolsep{\fill}}lrrrrrrrr@{}}
\toprule
Group & $n$ & JS & Pos./query & $\Delta$R@5 & $\Delta$R@10 & $\Delta$R@100 & $\Delta$N@10 & $\Delta$N@100\\
\midrule
\multicolumn{9}{@{}l}{\textbf{ESCI-US}: ResTD vs. Base-FT}\\
Low & 1,550 & 0.236 & 3.92 & +1.46 & +1.34 & +1.74 & +1.39 & +1.47\\
Medium & 1,550 & 0.394 & 4.48 & +1.82 & +1.62 & +2.17 & +1.95 & +2.14\\
High & 1,549 & 0.538 & 5.13 & +2.24 & +2.04 & +2.59 & +2.43 & +2.55\\
\addlinespace[2pt]
\multicolumn{9}{@{}l}{\textbf{ESCI-US}: ResTD vs. RBD (\textit{H}=1)}\\
Low & 1,550 & 0.236 & 3.92 & +1.04 & +0.93 & +1.17 & +0.85 & +1.04\\
Medium & 1,550 & 0.394 & 4.48 & +1.38 & +1.20 & +1.50 & +1.34 & +1.54\\
High & 1,549 & 0.538 & 5.13 & +1.70 & +1.51 & +1.87 & +1.82 & +1.88\\
\addlinespace[2pt]
\multicolumn{9}{@{}l}{\textbf{ESCI-ES}: ResTD vs. Base-FT}\\
Low & 450 & 0.432 & 3.79 & +0.58 & +0.86 & +0.79 & +1.34 & +1.35\\
Medium & 449 & 0.580 & 4.71 & +0.96 & +1.43 & +1.33 & +2.08 & +1.81\\
High & 449 & 0.658 & 5.16 & +0.81 & +1.43 & +1.43 & +1.75 & +1.72\\
\addlinespace[2pt]
\multicolumn{9}{@{}l}{\textbf{ESCI-ES}: ResTD vs. RBD (\textit{H}=1)}\\
Low & 450 & 0.432 & 3.79 & +0.46 & +0.59 & +0.51 & +0.95 & +0.97\\
Medium & 449 & 0.580 & 4.71 & +0.72 & +1.08 & +1.12 & +1.56 & +1.43\\
High & 449 & 0.658 & 5.16 & +0.65 & +1.13 & +1.17 & +1.30 & +1.41\\
\addlinespace[2pt]
\multicolumn{9}{@{}l}{\textbf{ESCI-JP}: ResTD vs. Base-FT}\\
Low & 512 & 0.409 & 3.93 & +0.95 & +1.04 & +1.20 & +1.40 & +1.39\\
Medium & 511 & 0.570 & 4.77 & +1.51 & +2.21 & +2.15 & +2.72 & +2.82\\
High & 511 & 0.652 & 5.13 & +2.47 & +2.55 & +2.76 & +3.37 & +3.09\\
\addlinespace[2pt]
\multicolumn{9}{@{}l}{\textbf{ESCI-JP}: ResTD vs. RBD (\textit{H}=1)}\\
Low & 512 & 0.409 & 3.93 & +0.71 & +0.76 & +0.83 & +0.83 & +1.00\\
Medium & 511 & 0.570 & 4.77 & +1.21 & +1.60 & +1.59 & +2.01 & +2.17\\
High & 511 & 0.652 & 5.13 & +1.95 & +1.98 & +2.08 & +2.64 & +2.35\\
\bottomrule
\end{tabular*}
\end{table}

%% file: tables/table_adjusted_retrieval_associations.tex
\begin{table}[!htb]
\centering\appendixtablestyle
\caption{Teacher disagreement and ResTD gains among multi-positive queries. Raw $\rho$ is the Spearman correlation. Adjusted coefficients are changes in gain (pp) per 0.1-nat increase in JS, controlling for $\log(\textnormal{1}+\text{positive items})$ and the corresponding score of the common model before fine-tuning
(Equation~\ref{eq:initial-query-score}). Brackets contain 95\% HC3 confidence intervals.}
\label{tab:adjusted-retrieval-associations}
\begin{tabular*}{\linewidth}{@{\extracolsep{\fill}}llrrrr@{}}
\toprule
 & & \multicolumn{2}{c}{Recall@100} & \multicolumn{2}{c}{NDCG@10}\\
\cmidrule(lr){3-4}\cmidrule(lr){5-6}
Locale & Compared with & Raw $\rho$ & Adjusted [95\% CI] & Raw $\rho$ & Adjusted [95\% CI]\\
\midrule
US & Base-FT & +0.080 & +0.16\,[+0.04,\,+0.28] & +0.090 & +0.19\,[+0.08,\,+0.30]\\
US & RBD (\textit{H}=1) & +0.081 & +0.10\,[+0.01,\,+0.19] & +0.105 & +0.17\,[+0.08,\,+0.25]\\
\addlinespace
ES & Base-FT & +0.060 & +0.16\,[\textminus{}0.15,\,+0.48] & +0.051 & +0.12\,[\textminus{}0.20,\,+0.43]\\
ES & RBD (\textit{H}=1) & +0.073 & +0.16\,[\textminus{}0.09,\,+0.40] & +0.055 & +0.06\,[\textminus{}0.19,\,+0.31]\\
\addlinespace
JP & Base-FT & +0.114 & +0.55\,[+0.26,\,+0.83] & +0.148 & +0.72\,[+0.42,\,+1.02]\\
JP & RBD (\textit{H}=1) & +0.123 & +0.43\,[+0.20,\,+0.65] & +0.168 & +0.62\,[+0.39,\,+0.85]\\
\bottomrule
\end{tabular*}
\end{table}

%% file: sections/appendix_collision_teacher_analysis.tex
\subsection{Collision-Aware Teacher Analysis}
\label{app:collision-teacher-analysis}

\subsubsection{Frequency and Magnitude of Teacher--SID Conflicts}
\label{app:collision-conflict-statistics}

For each locale and quantization level,
Table~\ref{tab:collision-conflict-statistics} reports two rates over all items.
\emph{SID edited} is the fraction whose stored code differs from the original
nearest-code assignment after collision resolution. \emph{Raw conflict} is
the fraction whose uncorrected teacher does not rank the stored code highest,
with residuals reconstructed along the final SID path
(Section~\ref{sec:collision}).

For each conflicting target, we compute the minimum one-hot mixing weight
needed to satisfy the margin constraint in Equation~\ref{eq:collision-minimal}
with $\delta=0.001$ for the corrected teacher defined in
Equation~\ref{eq:collision}:
\begin{equation}
  \epsilon^\star
  = \left[\frac{q_{\max}-q_y+\delta}{1-q_y+q_{\max}}\right]_0^1,
  \label{eq:minimal-collision-correction}
\end{equation}
where $q_y$ is the raw probability of the stored code and $q_{\max}$ is
the largest probability among competing codes. The applied mixing weight
is $\max\{\epsilon,\epsilon^\star\}$, with floor $\epsilon$.
The $\epsilon^\star$ summaries in
Table~\ref{tab:collision-conflict-statistics} are conditional on a raw conflict.

\input{tables/table_collision_conflict_statistics}

Conflicts become more frequent at greater quantization depth. At the first
level, conflict rates equal the rates of direct SID modification: 0.2\%
on US, 0.3\% on ES, and 0.2\% on JP. At this level, the residual has not
been affected by earlier collision-resolved assignments.
At the fourth level, conflict rates reach 10.1\%, 12.9\%, and 11.7\%,
exceeding the direct edit rates of 7.8\%, 9.4\%, and 8.8\%, respectively.
Changes to earlier codes alter the residuals used by subsequent teachers,
allowing conflicts even at positions whose codes were not modified.
Most conflicts require only a small correction:
96.3--97.1\% have $\epsilon^\star<0.1$ at the first level. The required
mixing weight increases at later levels, but 63.1--67.5\% of fourth-level
conflicts still fall below 0.1.

\input{tables/table_collision_correction_sensitivity}
\subsubsection{Retrieval Sensitivity to Collision Correction}
\label{app:collision-correction-sensitivity}

Table~\ref{tab:collision-correction-sensitivity} varies only teacher
correction in ResTD ($H=4$), holding the index, training, evaluation,
and three paired seeds fixed. \emph{None (raw teacher)} uses $\gQ$
without correction; \emph{Adaptive only} applies $\epsilon^\star$ with
zero floor. The remaining conditions use $\max\{\epsilon,\epsilon^\star\}$
with floors 0.05, 0.10, and 0.20. All corrected conditions use margin
$\delta=0.001$; the main experiments use $\epsilon=0.10$.

Relative to the raw teacher, correction with $\epsilon=0.10$ improves Recall@100 by 0.38,
0.34, and 0.43 percentage points on US, ES, and JP, respectively;
the corresponding NDCG@10 gains are 0.33, 0.43, and 0.50 points.
Adaptive-only correction recovers most of these gains; mean scores vary by
only 0.04--0.06 points across floors 0, 0.05, and 0.10 for each
locale--metric pair. Raising the floor to 0.20 reduces scores by
0.07--0.11 points relative to $\epsilon=0.10$. Stronger one-hot mixing
shifts probability mass from alternative codewords to the stored code.

%% file: tables/table_collision_conflict_statistics.tex
\begin{table}[!htbp]
\centering\appendixtablestyle
\caption{Frequency and magnitude of teacher--SID conflicts across ESCI locales and quantization levels. Rates use all items; $\epsilon^\star$ statistics are computed only over conflicting targets.}
\label{tab:collision-conflict-statistics}
\begin{tabular*}{\linewidth}{@{\extracolsep{\fill}}lrrrrrrrr@{}}
\toprule
 & & & & \multicolumn{5}{c}{Among conflicting targets}\\
\cmidrule(lr){5-9}
Locale & Level & \shortstack{SID edited\\(\%)} & \shortstack{Raw conflict\\(\%)} & Mean $\epsilon^\star$ & Median & P90 & P95 & \shortstack{$\epsilon^\star<\textnormal{0.1}$\\(\%)}\\
\midrule
US & 1 & 0.2 & 0.2 & 0.031 & 0.024 & 0.066 & 0.082 & 97.1\\
US & 2 & 0.6 & 0.9 & 0.039 & 0.030 & 0.081 & 0.101 & 94.8\\
US & 3 & 2.1 & 2.6 & 0.055 & 0.043 & 0.112 & 0.137 & 86.4\\
US & 4 & 7.8 & 10.1 & 0.083 & 0.071 & 0.157 & 0.201 & 67.5\\
\addlinespace
ES & 1 & 0.3 & 0.3 & 0.034 & 0.026 & 0.070 & 0.088 & 96.3\\
ES & 2 & 0.8 & 1.2 & 0.042 & 0.033 & 0.087 & 0.109 & 92.7\\
ES & 3 & 2.7 & 3.7 & 0.059 & 0.047 & 0.120 & 0.149 & 83.6\\
ES & 4 & 9.4 & 12.9 & 0.088 & 0.075 & 0.166 & 0.214 & 63.1\\
\addlinespace
JP & 1 & 0.2 & 0.2 & 0.033 & 0.025 & 0.069 & 0.085 & 96.8\\
JP & 2 & 0.7 & 0.8 & 0.041 & 0.032 & 0.085 & 0.106 & 93.4\\
JP & 3 & 2.5 & 3.2 & 0.058 & 0.046 & 0.118 & 0.145 & 84.2\\
JP & 4 & 8.8 & 11.7  & 0.086 & 0.073 & 0.162 & 0.207 & 65.4\\
\bottomrule
\end{tabular*}
\end{table}

%% file: tables/table_collision_correction_sensitivity.tex
\begin{table}[H]
\centering\appendixtablestyle
\caption{Retrieval performance (\%) under collision-aware teacher correction for ResTD (\textit{H}=4) on ESCI. Values are mean \textpm{} standard deviation across three paired training seeds in each locale.}
\label{tab:collision-correction-sensitivity}
\begin{tabular*}{\linewidth}{@{\extracolsep{\fill}}lrrrrrrr@{}}
\toprule
 & & \multicolumn{2}{c}{ESCI-US} & \multicolumn{2}{c}{ESCI-ES} & \multicolumn{2}{c}{ESCI-JP}\\
\cmidrule(lr){3-4}\cmidrule(lr){5-6}\cmidrule(lr){7-8}
Correction strategy & Floor $\epsilon$ & R@100 & N@10 & R@100 & N@10 & R@100 & N@10\\
\midrule
None (raw teacher) & -- & \appendixstat{39.20}{0.13} & \appendixstat{12.66}{0.08} & \appendixstat{35.04}{0.12} & \appendixstat{13.79}{0.10} & \appendixstat{32.43}{0.14} & \appendixstat{13.68}{0.11}\\
Adaptive only & 0 & \appendixstat{39.53}{0.13} & \appendixstat{12.95}{0.08} & \appendixstat{35.34}{0.12} & \appendixstat{14.17}{0.09} & \appendixstat{32.81}{0.13} & \appendixstat{14.12}{0.11}\\
Adaptive + floor & 0.05 & \appendixstat{39.54}{0.12} & \appendixstat{12.95}{0.08} & \appendixstat{35.36}{0.12} & \appendixstat{14.21}{0.09} & \appendixstat{32.85}{0.13} & \appendixstat{14.16}{0.10}\\
\rowcolor{mainresultrow}\textbf{Adaptive + floor} & \textbf{0.10} & \appendixbeststat{39.58}{0.11} & \appendixbeststat{12.99}{0.07} & \appendixbeststat{35.38}{0.13} & \appendixbeststat{14.22}{0.07} & \appendixbeststat{32.86}{0.09} & \appendixbeststat{14.18}{0.08}\\
Adaptive + floor & 0.20 & \appendixstat{39.51}{0.13} & \appendixstat{12.92}{0.09} & \appendixstat{35.31}{0.12} & \appendixstat{14.14}{0.10} & \appendixstat{32.77}{0.14} & \appendixstat{14.07}{0.11}\\
\bottomrule
\end{tabular*}
\end{table}

%% file: sections/appendix_hyperparameter_sensitivity.tex
\subsection{Hyperparameter Sensitivity}
\label{app:hyperparameter-sensitivity}

\begin{figure}[!tbp]
  \centering
  \includegraphics[width=0.82\linewidth,trim=0 0 173bp 0,clip]{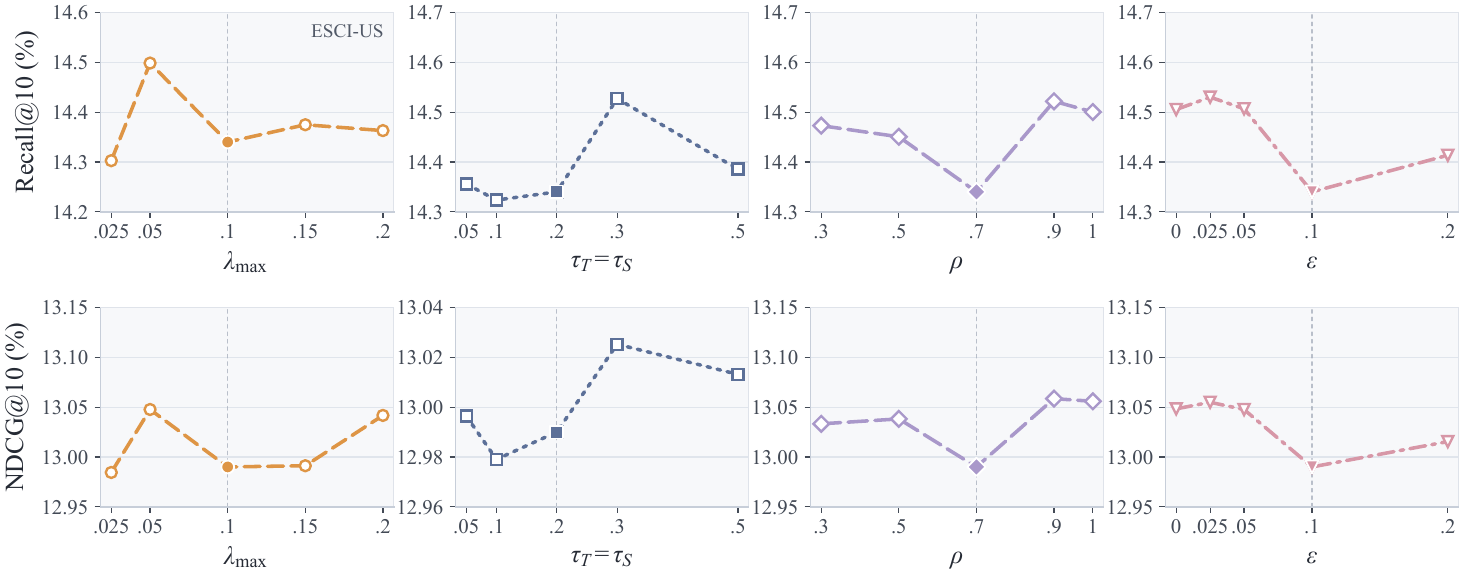}\par\vspace{4pt}
  \includegraphics[width=0.82\linewidth,trim=0 0 173bp 0,clip]{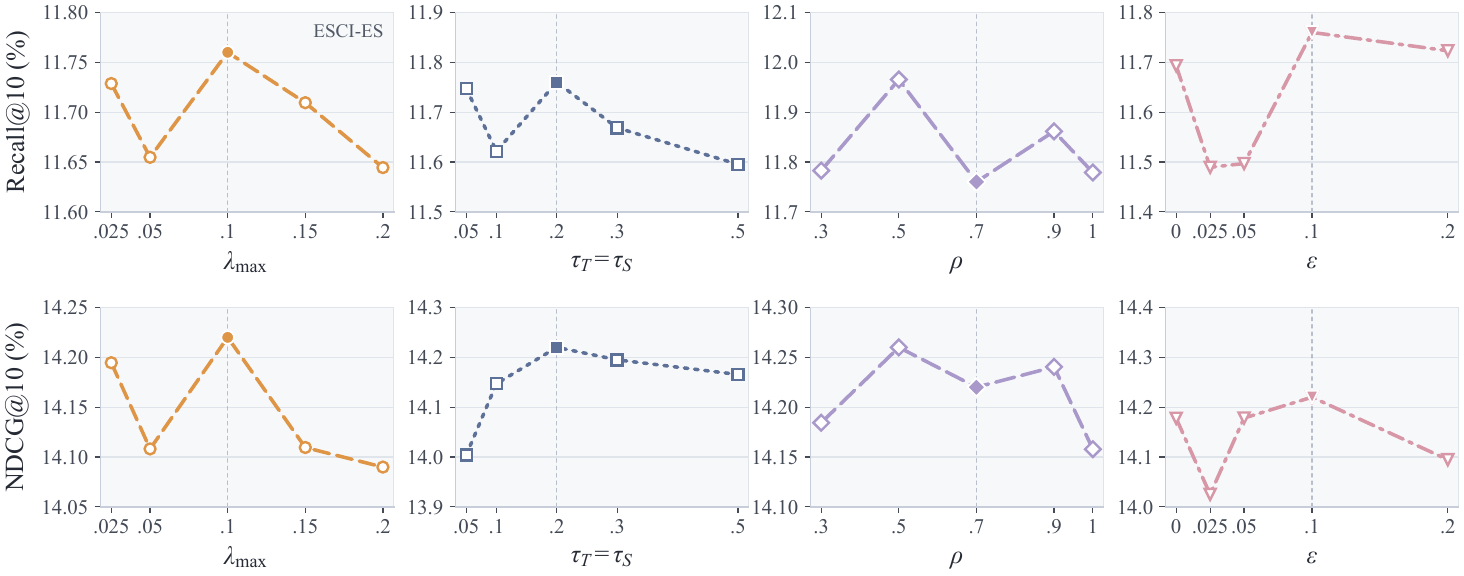}\par\vspace{4pt}
  \includegraphics[width=0.82\linewidth,trim=0 0 173bp 0,clip]{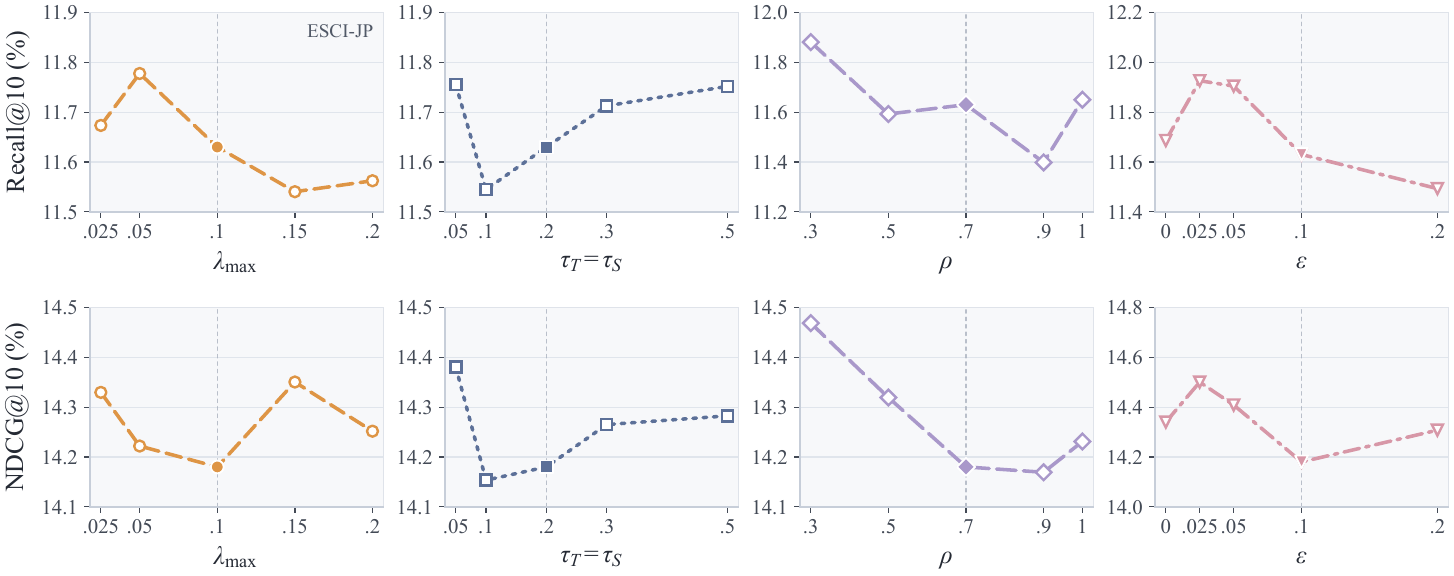}
  \caption{Hyperparameter sensitivity on ESCI-US (top), ESCI-ES (middle),
  and ESCI-JP (bottom). Rows show Recall@10 and NDCG@10;
  columns vary $\lambda_{\max}$, $\tau_T=\tau_S$, and $\rho$ one at a time.
  Dashed lines and filled markers indicate the shared defaults.
  Each panel uses its own vertical scale.}
  \label{fig:sensitivity-us}
  \label{fig:sensitivity-es}
  \label{fig:sensitivity-jp}
\end{figure}

We vary $\lambda_{\max}\in\{0.025,0.05,0.1,0.15,0.2\}$,
$\tau_T=\tau_S\in\{0.05,0.1,0.2,0.3,0.5\}$, and
$\rho\in\{0.3,0.5,0.7,0.9,1.0\}$ one at a time, keeping the remaining
settings fixed. All locales use the defaults in Section~\ref{sec:exp-setup}.
Figure~\ref{fig:sensitivity-us} reports Recall@10 and NDCG@10 for each
locale.
Appendix~\ref{app:collision-correction-sensitivity} separately evaluates
the collision-correction floor $\epsilon$ over three training seeds.

\paragraph{Auxiliary loss weight.}
Increasing $\lambda_{\max}$ does not yield a monotonic improvement.
Among the tested values, $0.05$ gives the highest Recall@10 on US and JP,
while ES peaks at the default $0.1$. On JP, the two metrics favor
different weights, with NDCG@10 highest at $0.15$.
The coefficient in Equation~\ref{eq:total} is also subject to the
5\% SID-loss cap. Consequently, increasing $\lambda_{\max}$ need not
produce a proportional increase in the effective auxiliary contribution.

\paragraph{Teacher and student temperature.}
The preferred temperature varies across locales: both metrics peak at
$0.3$ on US, $0.2$ on ES, and $0.05$ on JP within the tested grid.
The effect is not monotonic, and the common default $0.2$ does not
maximize every curve. Temperature changes the concentration of the
codeword distributions and the gradient scale of the student objective
(Equation~\ref{eq:rbd-geometry}). Since $\tau_T$ and $\tau_S$ are varied
jointly, this sweep measures their combined effect and does not isolate
teacher softness from student optimization.

\paragraph{Horizon discount.}
Larger $\rho$ assigns more relative weight to distant targets in
Equation~\ref{eq:retrace}. The largest plotted scores occur at
$\rho=0.9$ on US, $0.5$ on ES, and $0.3$ on JP.
For $H=L=4$ with all positions valid, the total weight at offset $s$ is
proportional to $(4-s)\rho^s$. Thus, $\rho=1$ gives offset weights in
the ratio $4{:}3{:}2{:}1$, because fewer state--target pairs are available
at longer offsets. All tested discounts retain supervision at all four
offsets, so this comparison concerns the allocation of trajectory
supervision with its horizon range fixed.

The shared defaults specify a common configuration across locales.
These sweeps describe sensitivity to individual parameters with the other
settings fixed; interactions among parameters are not evaluated.

%% file: sections/appendix_case_study.tex
\Needspace{48\baselineskip}
\section{Qualitative Case Study}
\label{app:case-study}

\begin{figure}[H]
  \centering
  \includegraphics[width=\linewidth]{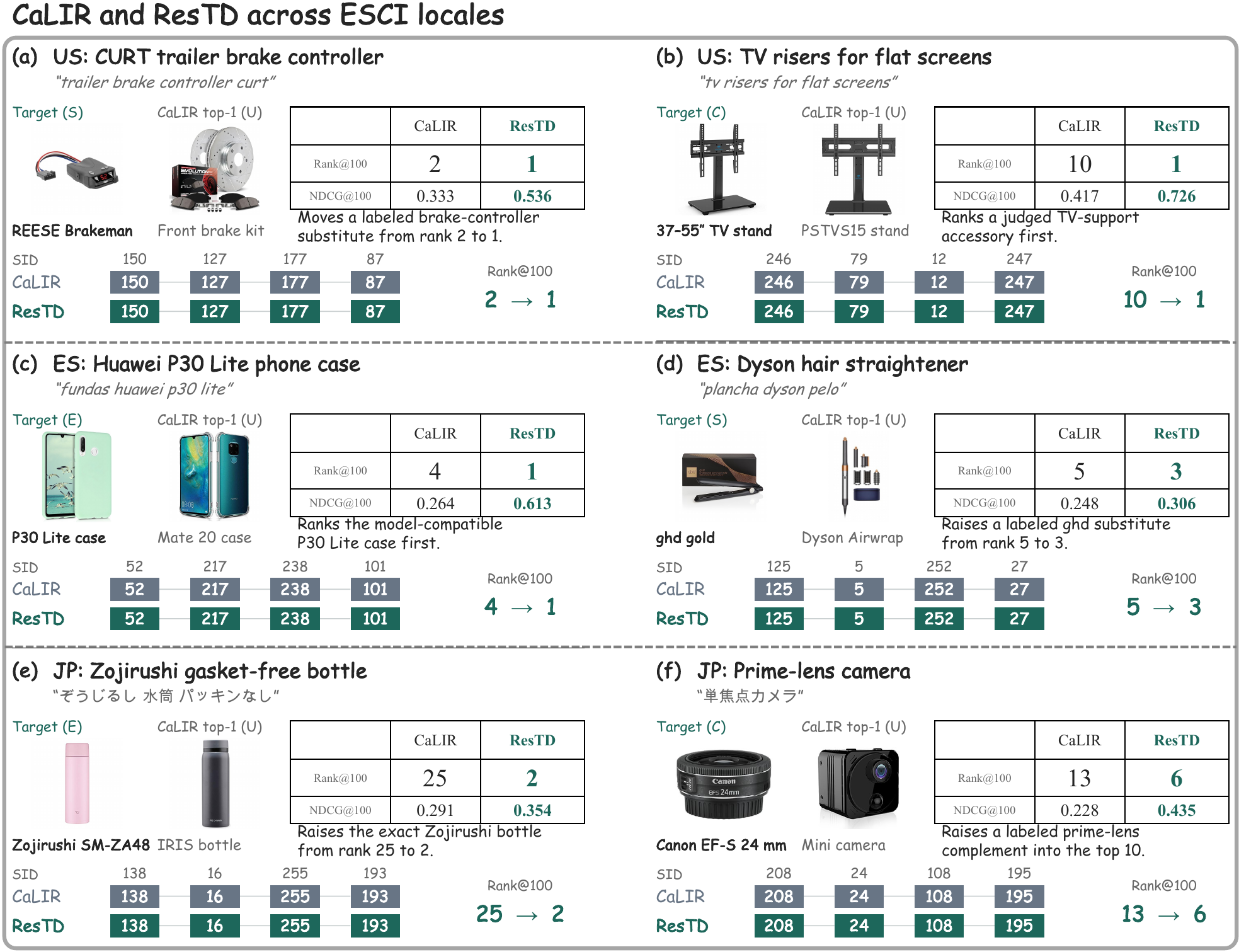}
  \caption{Retrieval examples from ESCI-US, ES, and JP. Each panel shows
  a judged target and CaLIR's top-ranked product, with target SIDs, ranks,
  and query-level NDCG@100 for both models.}
  \label{fig:case-study}
\end{figure}

Figure~\ref{fig:case-study} presents six selected ESCI queries across US,
ES, and JP, covering Exact (E), Substitute (S), and Complement (C)
judgments. Each panel pairs a judged target with CaLIR's top-ranked
product. U denotes an unjudged product, for which the dataset provides
no relevance label. Both models use the same SID index and a beam width
of 100. Rank@100 is the target SID's position among the top 100 results,
and NDCG@100 measures query-level ranking quality on the $[0,1]$ scale.
The SID rows show the complete target sequence retrieved by each model.
The exact-match cases involve specific compatibility and product-attribute
requirements. In panel (c), ResTD promotes the Huawei P30 Lite case from
rank 4 to 1; CaLIR places a case for the Mate 20 first. In panel (e),
the query specifies both the Zojirushi brand and a gasket-free bottle
design, and the exact-match target rises from rank 25 to 2.
These examples require distinguishing products within the same category:
phone-model compatibility determines which case fits, while the bottle
query combines a brand preference with a design requirement, both of
which must be met for a retrieved product to satisfy the query in full.

ResTD also promotes judged substitutes and complements. The substitute
brake controller in panel (a) rises from rank 2 to 1, and the ghd straightener
in panel (d) advances from rank 5 to 3 for the Dyson query. Among
complementary products, the TV stand in panel (b) moves from rank 10 to 1,
and the camera lens in panel (f) enters the top ten at rank 6, up from rank 13.
The relevance judgments in these panels extend beyond exact product
matches to alternatives and accessories associated with the query.
Both models retrieve the complete target SID in all six cases, so the
observed differences concern the ranking of targets already present in
the retrieved lists. NDCG@100 also increases for every query. The gain
depends on the full ranking: the bottle target moves from rank 25 to 2
while NDCG@100 rises from 0.291 to 0.354; the phone case moves from rank
4 to 1 while NDCG@100 rises from 0.264 to 0.613. Target rank tracks one
judged product, whereas NDCG@100 also reflects the positions and
binary relevance of the other retrieved products within the top 100 results.

